\documentclass[sort&compress,AMA,STIX1COL]{WileyNJD-v2}
\articletype{Research Article}%

\received{26 March 2022}
\revised{- June 2022}
\accepted{- June 2022}

\begin{document}

\title{\textcolor{black}{Unknown Piecewise Constant Parameters Identification with Exponential Rate of Convergence}\protect}

\author[1]{Anton Glushchenko*}

\author[1]{Konstantin Lastochkin}

\authormark{Glushchenko A. \textsc{et al}}

\address[1]{\orgdiv{Ya.Z. Tsypkin Laboratory of Adaptive and Robust Systems}, \orgname{V.A. Trapeznikov Institute of Control Sciences of Russian Academy of Sciences}, \orgaddress{\state{Moscow}, \country{Russia}}}

\corres{*Anton Glushchenko, Russia, Moscow, Profsoyuznaya street, h.65, 117997, ICS RAS. \email{aiglush@ipu.ru}}

\abstract[Summary]{\textcolor{black}{The scope of this research is the identification of unknown piecewise constant parameters of linear regression equation under the finite excitation condition. Compared to the known methods, to make the computational burden lower, only one model to identify all switching states of the regression is used in the developed procedure with the following two-fold contribution. First of all, we propose a new truly online estimation algorithm based on a well-known DREM approach to detect switching time and preserve time alertness with adjustable detection delay. Secondly, despite the fact that a switching signal function is unknown, the adaptive law is derived that provides global exponential convergence of the regression parameters to their true values in case the regressor is finitely exciting somewhere inside the time interval between two consecutive parameters switches. The robustness of the proposed identification procedure to the influence of external disturbances is analytically proved. Its effectiveness is demonstrated via numerical experiments, in which both abstract regressions and a second-order plant model are used.}}

\keywords{identification, linear regression, piecewise constant parameters, switching, finite excitation, monotonicity, extension and mixing}

\jnlcitation{\cname{%
\author{Glushchenko A.}, and
\author{K. Lastochkin}} (\cyear{2022}), 
\ctitle{\textcolor{black}{Unknown Piecewise Constant Parameters Identification with Exponential Rate of Convergence}}, \cjournal{International Journal of Adaptive Control and Signal Processing}, \cvol{2022;00:1--31}.}

\maketitle

\section{Introduction}\label{sec1}

In recent years, the problem of identification of regression equation (RE) unknown parameters has received considerable attention from the scientific community in the field of control theory \cite{c1}. The mentioned problem is of high importance as, using special parameterizations \cite{c2}, most of the adaptive control theory problems can be reduced to the identification of the RE unknown parameters. For example, in \cite{c3, c4} a parametrization is proposed, in which the plant dynamics identification problem is converted into the estimation of the unknown initial conditions. In \cite{c5} an approach is developed to represent the classical problems of the model reference adaptive control as the identification of unknown ideal parameters of a control law that is written as RE. The known major studies related to the solution of the RE parameters identification problem, in general, can be divided into two main groups with the following objectives:
\begin{itemize}
\item ensure exponential convergence of the parameter error to zero under weak regressor excitation requirements \cite{c1,
c6,c7,c8,c9,c10,c11,c12,c13,c14,c15,c16,c17,c18,c19,c20,c21,c22,c23};

\item improve the quality of identification for RE with over-parametrization \cite{c1, c17, c24, c25, c26, c27}.
\end{itemize}

The present work is devoted to the development of methods, which belong to the first group.

It is well known \cite{c1, c2} that the classical gradient-descent and least-squares-based adaptive laws provide exponential convergence of the parameter error to zero only if the strict requirement of the regressor persistent excitation (PE) is met. This means that the regressor is to be sufficiently rich \cite{c28}, i.e. it is expected to contain as many different spectral lines as the number of RE unknown parameters is. But that is not the truth for many practical problems as far as the normal functioning mode of most plants is concerned. Therefore, in order to obtain the true values of the unknown parameters, it becomes necessary to artificially provide the required regressor richness by injection of test signals into the plant control input. For example, the procedures \cite{c29, c30, c31} can be used for that. On the other hand, in many applications such injection can lead to damage of actuators, energy consumption increase, and cause situations when the plant functions out of standard operating mode.

As a result, new adaptive laws have been proposed in the literature \cite{c6, c7, c8, c9, c10, c11, c12, c13, c14, c15, c16} to guarantee exponential convergence of the unknown parameters estimates to their true values in case of the regressor initial (IE) or finite (FE) excitation. Most of these approaches are based on the composite/combined adaptive laws that use previously stored and current values of the regression function and the regressor concurrently. Some off-line algorithms are applied to store data in a special data stack and ensure that they are sufficiently rich \cite{c6, c15}. Thus, using some kind of memory, such laws provide a parameter error exponential convergence to zero even when the regressor excitation has already vanished. Another way to provide exponential convergence is to filter the RE using different filters with the integral or strong inertial properties \cite{c7, c8, c9, c10, c11, c12, c13, c14}. Also, recently the energy pumping-and-damping injection principle \cite{c32} has been successfully applied to generate the PE regressor from the IE one \cite{c18, c19}. In contrast to the above-considered adaptive laws that require the data stack processing, this approach does not use offline operations, but, like the schemes \cite{c7, c8, c9, c10, c11, c12, c13, c14}, has strong inertial properties. In \cite{c20, c21, c22, c23} adaptive laws are proposed, which, unlike the previously considered ones, provide finite time identification of the RE unknown parameters if IE or FE condition is met. A more detailed review of some modern methods to relax the PE condition can be found in \cite{c1}.

However, the laws \cite{c6, c7, c8, c9, c10, c11, c12, c13, c14, c15, c16, c17, c18, c19, c20, c21, c22, c23} have been developed only to solve the identification problems of the time-invariant parameters. Their applicability to estimate the piecewise constant (PC) parameters is usually tested via numerical experiments only and, with a number of exceptions like \cite{c13, c14, c20}, does not have a sufficiently rigorous analytical proof.

Adaptive laws \cite{c6, c7, c8, c9, c10, c11, c12, c13, c14, c15, c16, c17, c18, c19, c20, c21, c22} have difficulties in solving problems of piecewise constant unknown parameters identification due to imperfect data storage/filtering procedures and inertial characteristics of filters, which are used for the RE processing. The point is that when the data on RE with not time-invariant parameters are stored in the data stack or the regression with piecewise constant parameters is filtered with integral-based filters, then the combined/composite adaptive laws lose their ability to track true values of the unknown parameters. The reason is the superpositional mixing of data on the regressions with different values of parameters. In this case only estimations boundedness could be guaranteed \cite{c13, c14, c33}. A more detailed analytical and numerical comparison of some modern combined adaptive laws to solve the problem of piecewise constant unknown parameters identification is given in \cite{c33} (see Fig.4).

Many authors \cite{c13, c14, c18, c20, c33, c34, c35, c36} have pointed out the vital necessity to derive adaptive laws that could also provide the identification of PC parameters with exponential or finite-time convergence. In \cite{c18}, it is mentioned that a reinitialization procedure is required for a scheme with the excited regressor generation. It should be applied each time when the unknown parameters of RE change their values. Various methods\cite{c34, c35, c36} are proposed to overcome superpositional mixing of data in the data stack caused by the unknown parameters switch. \textcolor{black}{In \cite{R1} the identification problem of piecewise constant change rate of unknown parameters is considered.} In \cite{c37}, the dynamic regressor extension and mixing scheme \cite{c17} sensitivity to the unknown parameters switching is discussed. The study \cite{c20} is devoted to the question of the applicability of the adaptive law with finite time convergence to identify the unknown PC parameters. Thus, the problem of identification of RE with switched parameters is relevant and deserves a stand-alone detailed discussion.

In the machine learning theory, the classification methods \cite{c38}, principal component analysis \cite{c39}, ARX (SNARX) regression models \cite{c40} are well known to be able to find the unknown switched parameters. Such models are trained offline or in discrete time on the basis of a data array that includes the measured values of the regressor and function at some time instants. However, the application of these and other machine-learning-based techniques to solve the identification and adaptive control problems online in continuous time faces difficulties, which are mainly caused by high computational costs in the case of an increasing number of measurements and higher system dimensions.

On the other hand, the problem of PC unknown parameters identification is well-developed as a part of the switched systems adaptive control theory \cite{c41}. The general formulation of the identification problem for switched systems includes: (1) the identification of a discrete switching function, (2) the choice of the current model of the system from a known set or a generation of a new one, (3) the estimation of the current values of the unknown PC parameters of the system (usually they include the state and control matrices of the piecewise-linear or piecewise-affine plant).

\textcolor{black}{In \cite{c42, c43} the Lyapunov-like laws of unknown PC parameters identification of piecewise-linear or piecewise-affine plants are proposed. Separate regression model with adjustable parameters and related adaptive law are introduced for each possible set of the plant unknown parameters values. The switching signal function between such models is considered to be known. At a certain time instant only the parameters of the active model are adjusted. The result of identification of the plant unknown PC parameters is a piecewise-continuous function that switches between estimations of parameters of each separate model at known time instants of the plant parameters change. The disadvantages of \cite{c42, c43} are as follows: (1) the unknown parameters identification error converges only under PE condition, (2) only the active model parameters are adjusted, (3) the switching signal function is to be known {\it a priori}, (4) an arbitrarily large number of identification laws are used instead of one. In order to relax PE condition and adjust the parameters of inactive models, identification laws are proposed in \cite{c35, c44}, which are based on the concurrent learning. After another plant parameters switch, information about the current unknown parameters is added to a data stack of a currently active model in the course of the regressor finite excitation time range. Owing to this technique, the requirement of the regressor persistent excitation is relaxed and the parameters are adjusted even for the inactive models (in case they have been active at least once over the interval of the regressor finite excitation). However, concurrent-learning-based adaptive laws \cite{c35, c44}: (1) are applicable only if the switching signal function is known {\it a priori}, (2) use an arbitrarily large number of identification laws, and (3) require to store and process large amounts of data on each adjustable model.}

In a recent paper \cite{c45}, an approach based on a combination of earlier results \cite{c15, c46, c47, c48} has been proposed to solve the switched system identification problem without the knowledge of both the unknown parameters switching time instants and function to define what model to make active at the moment.

In order to identify the function that defines the switching time instants, the above-mentioned research proposes to compare the regression equations stored in the data stack with the current regression formed using the measured data from the plant. Such a comparison is based on the data-driven projection subspace method \cite{c46}. Using the principles from \cite{c46}, a residual signal is generated and used as an indicator of the system unknown parameters change (switch). Then, applying the numerical robust algorithm of residual analysis \cite{c47, c48}, the switching signal function is estimated with almost arbitrary accuracy.

After that the residual value is calculated for each model, the information about which is stored in the data stack. If such value is less than the preset threshold for some model, then it is assumed that the plant is described by such model. And it is marked as ‘active’. If no model with the residual below the threshold has been found, then a new one is generated. A data stack is formed and processed for each model that was active for at least one time. This relaxes excitation conditions and makes it possible to identify the parameters of each model all the time, even when the model is not active at the moment.

To identify the parameters of a switched system, according to \cite{c45}, the number of the adaptive laws to be introduced coincide with the number of detected models. And, as the estimation of the unknown parameters is performed on basis of the data from the data stack, this method requires only a finite excitation of the regressor over the time range when the model is active to provide exponential convergence of the parameter error to zero \cite{c15}.

Thus, the approach from \cite{c45} ensures the PC unknown parameters tracking without knowledge about switching signal function, which is its main advantage over other procedures \cite{c35, c42, c43, c44}. However, over time, the number of models used in \cite{c45}can increase indefinitely, and the necessity to adjust simultaneously all parameters of all models, as well as store, monitor and process a high number of data stacks, requires high computational power of the hardware in use. In addition, the presence of several models makes it difficult to analyze the stability of a closed-loop system, when such identification method is applied as a part of the adaptive control system. The problem is that the control law is discontinuous when several models are in use, and the proof of stability requires the application of a multiple Lyapunov function \cite{c41}.

In this regard, the actual problem is to develop a law of the unknown piecewise constant parameters estimation, which does not use several separate adaptive laws, as well as offline operations of data monitoring and stacking, but is able to identify piecewise constant unknown parameters and ensure the exponential rate of convergence under FE condition.

Thus, we propose a new procedure to track unknown PC parameters with the following contribution:

\textbf{C1)} on the basis of the dynamic regressor extension and mixing procedure \cite{c17}, a new online algorithm with adjustable detection lag is proposed to estimate switching time instants. \textcolor{black}{In contrast to the solution \cite{c45}, the proposed algorithm is truly online and does not involve an offline operations of data monitoring and stacking;}

\textbf{C2)} based on the proposed estimation algorithm, the novel adaptive law is derived. It ensures global exponential convergence of the unknown PC parameters estimates to their true values if the regressor is finitely exciting somewhere inside the time interval between two consecutive changes of the parameters. \textcolor{black}{Opposed to existing PE requirement relaxation methods \cite{c6, c7, c8, c9, c10, c11, c12, c13, c14, c15, c16, c17, c18, c19, c20, c21, c22, c23}, it is analytically proven that the proposed method provides the exponential convergence rate of the unknown piecewise constant parameters identification under the condition that the switching signal function is unknown. In contrast to \cite{c35, c42, c43, c44, c45}, only one adaptive law is used to solve the identification problem.}

To the best of the authors’ knowledge, the proposed method of piecewise constant unknown parameters identification is the first solution that ensures all the above-stated properties simultaneously.

The further part of the manuscript is arranged as follows. Section 2 gives a rigorous mathematical problem statement, Section 3 is to propose (1) an algorithm to estimate the switching time instants, and (2) an adaptive law that provides exponential convergence of the error of piecewise constant unknown parameters estimation. Section 4 presents the discussion that includes tuning guidelines and comparison analysis. Section 5 presents the results of the numerical experiments.

The following notation is used throughout the paper. $\mathbb{R}^n$ and $\mathbb{R}^{n \times m}$ denote the sets of $n$-dimensional real vectors and $n \times m$-dimensional real matrices respectively, $|.|$ represents the absolute value, $\|.\|$ denotes Euclidean norm of a vector, the identity and nullity $n \times m$-dimensional matrices are denoted as $I_{n \times m}$ and $0_{n \times m}$ respectively. $E \{.\}$ is the operator to calculate the mean value, $var \left(.\right)$ is the operator to calculate the variance. $det\{.\}$ stands for a matrix determinant, $adj\{.\}$ -- for an adjoint matrix, $L_{\infty}$ is the space of all essentially bounded functions. We also use the fact that for all (possibly singular) ${n \times n}$ matrices $M$ the following holds: $adj \{M\} M = det \{M\}I_{n \times n}$. $O(.)$ stands for the 'Big $O$' notation to estimate complexity of algorithms.

The following definition from \cite{c2, c6,c7,c8,c9,c10,c11,c12,c13,c14,c15,c16,c17} is introduced.

\begin{definition}\label{dfn1}
A regressor $\varphi \left(t\right) \in \mathbb{R}^n$ is finitely exciting ($\varphi \left(t\right)  \in$  FE) over the time range $\left[t_r^+;t_e\right]$ if there exist $t_e \ge t_r^+ \ge0$ and $\alpha  > 0$ such that the following inequality holds:
\begin{eqnarray}\label{eq1}
\int\limits_{t_r^+}^{t_e} {\varphi \left( \tau  \right){\varphi ^{\rm T}}\left( \tau  \right)d} \tau  \ge \alpha I_{n \times n},
\end{eqnarray}
where $\alpha$  is an excitation level.
\end{definition}


\section{Problem statement}\label{sec2}

The identification problem of unknown piecewise constant parameters of a linear regression equation is considered:
\begin{eqnarray}\label{eq2}
\forall t \ge t_0^ + {\rm{,}}\;\;y\left( t \right) = {\varphi ^{\rm{T}}}\left( t \right){\Theta _{\kappa \left( t \right)}}{\rm{,}}
\end{eqnarray}
where $\varphi \left( t \right) \in {\mathbb{R}^{n \times m}}$, $y\left( t \right) \in {\mathbb{R}^{m \times p}}$ are measurable regressor and function respectively, ${\Theta _{\kappa \left( t \right)}} \in {\mathbb{R}^{n \times p}}$ is a matrix of unknown piecewise constant parameters, $\kappa \left( t \right) \in \Xi  = \left\{ {1,2, \ldots ,N} \right\}$ is an unknown discrete function, which defines the time points when the regression parameters switch to their new values, $t_0^+$ is a known initial time instant, $N$ is a number of different regression switching states (sets of parameters ${\Theta _{\kappa \left( t \right)}}$). To be explicit, $\kappa \left( t \right)$ and  ${\Theta _{\kappa \left( t \right)}}$ are assumed to be right-continuous:
\begin{eqnarray}\label{eq3}
\forall t \ge t_0^ + \; \kappa \left( t \right) = \mathop {{\rm{lim}}}\limits_{\tau  \to t_i^ + } \kappa \left( \tau  \right), \; {\Theta _{\kappa \left( t \right)}} = \mathop {{\rm{lim}}}\limits_{\tau  \to t_i^ + } {\Theta _{\kappa \left( \tau  \right)}}.
\end{eqnarray}

In general case, the signal $\kappa \left( t \right)$ is used to represent the switching sequence:
\begin{eqnarray}\label{eq4}
\Sigma  = \left\{ {\left( {{j_0}{\rm{,\;}}t_0^ + } \right){\rm{,\;}}\left( {{j_1}{\rm{,\;}}t_1^ + } \right){\rm{\;}}{\rm{,}} \ldots {\rm{,}}\left( {{j_{i - 1}}{\rm{,\;}}t_{i - 1}^ + } \right){\rm{,}}\left. {\left( {{j_i}{\rm{,\;}}t_i^ + } \right){\rm{,}} \ldots } \right|{j_i} \in \Xi {\rm{,}}\;\;{j_i} \ne {j_{i + 1}}{\rm{,}}\;\;i \in \mathbb{N}} \right\}.
\end{eqnarray}

This means that $\forall t \in \left[ {t_i^ + {\rm{;}}\;t_{i + 1}^ + } \right){\rm{,}}\;\kappa \left( t \right) = {j_i}{\rm{,}\;}{\Theta _{\kappa \left( t \right)}} = {\Theta _{{j_i}}}$. For the sake of brevity, the element of the set $\Sigma$, which corresponds to $\left[ {t_i^ + {\rm{;}}\;t_{i + 1}^ + } \right)$, is denoted as ${\theta _i}$ $\left( {\forall t \in \left[ {t_i^ + {\rm{;}}\;t_{i + 1}^ + } \right){\rm{}}\;{\theta _i} = {\Theta _{\kappa \left( t \right)}} = {\Theta _{{j_i}}}} \right)$. So, the equation \eqref{eq2} is rewritten as follows:
\begin{eqnarray}\label{eq5}
\begin{array}{c}
\forall t \ge t_0^ + \; y\left( t \right) = {\varphi ^{\rm{T}}}\left( t \right)\theta \left( t \right){\rm{,}}\\
\theta \left( t \right) = {\theta _i} = {\theta _0} + \sum\limits_{q = 1}^i {\Delta _q^\theta h\left( {t - t_q^ + } \right)}, \;\dot \theta \left( t \right) = \sum\limits_{q = 1}^i {\Delta _q^\theta \delta \left( {t - t_q^ + } \right)} {\rm{,}}
\end{array}
\end{eqnarray}
where $ \Delta _q^\theta  = {\theta _i} - {\theta _{i - 1}} = {\Theta _{{j_i}}} - {\Theta _{{j_{i - 1}}}}$ is an amplitude of ${\theta _{i - 1}}$ value change at time point $ t_i^ + $, $ h\left( . \right)$ and $ \delta \left( . \right)$ are the Heaviside and Dirac functions respectively.

Additionally, the switching time instants are written as a time sequence:
\begin{eqnarray}\label{eq6}
\Im  = \left\{ {t_0^ + {\rm{,}}\;t_1^ + {\rm{}}\;{\rm{,}} \ldots {\rm{,}}\;t_{i - 1}^ + {\rm{,}}\;\left. {t_i^ + {\rm{,}} \ldots } \right|i \in \mathbb{N}} \right\}{\rm{.}}
\end{eqnarray}

The following assumption is introduced with respect to ${\theta _i}$, the time range $\left[ {t_i^ + {\rm{;}}\;t_{i + 1}^ + } \right)$ and the regressor $\varphi \left( t \right)$.

\textbf{Assumption 1.} Let $\exists {\Delta _\theta } > 0,\;{T_{{\rm{min}}}} > \mathop {\min }\limits_{\forall i \in \mathbb{N}} {T_i} > 0$ such that $\forall i \in \mathbb{N}$ simultaneously:
\begin{enumerate} 
  \item[1)] $t_{i + 1}^ +  - t_i^ +  \ge {T_{{\rm{min}}}}$, $\left\| {{\theta _i} - {\theta _{i - 1}}} \right\| = \left\| {\Delta _q^\theta } \right\| \le {\Delta _\theta }{\rm{;}}$
  \item[2)] $\varphi \left( t \right) \in {\rm{FE}}$ over the time range $\left[ {t_i^ + {\rm{;}}\;t_i^ +  + {T_i}} \right]$ with the excitation level ${\alpha _i}$;
  \item[3)] ${\varphi \left( t \right) \in {\rm{FE}}}$ over the time range $\left[ {\hat t_i^ + {\rm{;}\;}t_i^ +  + {T_i}} \right]$ with the excitation level ${\overline \alpha _i}$, where ${\alpha _i} > {\overline \alpha _i} > 0,{\rm{}}\;\hat t_i^ +  \in \left[ {t_i^ + {\rm{;}}\;t_i^ +  + {T_i}} \right)$;
  \item[4)] $\varphi \left( t \right) \in {L_\infty }$;
  \item[5)] any of the following conditions holds:
  \begin{enumerate} 
 \item[5.1)] $i \le {i_{{\rm{max}}}} < \infty {\rm{;}}$
 \item[5.2)] $\forall q \in \mathbb{N} \; \left\| {\Delta _q^\theta } \right\| \le {c_q}{e^{ - k\left( {t_q^ +  - t_0^ + } \right)}}, \; {c_q} > {c_{q + 1}}{\rm{,}}$ where $ k > 0$ is a known constant.
  \end{enumerate}
\end{enumerate}

The following goals are to be achieved under Assumption 1:
\begin{eqnarray}\label{eq7}
\tilde t_i^ +  \le {T_i}, \; \mathop {{\rm{lim}}}\limits_{t \to \infty } \left\| {\tilde \theta \left( t \right)} \right\| = 0 \; \left( {exp} \right){\rm{,}}
\end{eqnarray}
where $\hat t_i^ +, \; \tilde t_i^ +  = \hat t_i^ +  - t_i^ + $ are the estimate of ${i^{th}}$ element of the sequence \eqref{eq6} and the error of such estimation, $\hat \theta \left( t \right), \; \tilde \theta \left( t \right) = \hat \theta \left( t \right) - \theta \left( t \right)$ are the parameters estimates and error respectively.

According to the stated goal \eqref{eq7}, the adaptive laws for $\hat \theta \left( t \right)$ and $\hat t_i^ + $ are to be derived. They must ensure: \linebreak 1) the bounded value of the estimation error of each element of the sequence \eqref{eq6}, 2) the exponential convergence to zero of the parameters error $\tilde \theta \left( t \right)$.

\emph{Remark 1:} The first part of Assumption 1 requires a finite frequency and amplitude of the step change of the unknown parameters, which are classical requirements for the switched systems \cite{c41} and identification \cite{c2} theories respectively. The second and third parts of the assumption present a necessary and sufficient condition to identify the true values of all elements of the $i^{\rm th}$ matrix of unknown parameters \cite{c49}. The fourth part of the assumption can be satisfied \cite{c2} by the multiplication of the regression \eqref{eq2} with the normalizing coefficient ${n_s}\left( t \right) = {\textstyle{1 \over {1 + {\varphi ^{\rm{T}}}\left( t \right)\varphi \left( t \right)}}}$. The fifth part of the assumption requires boundedness of \linebreak $\left\| {\theta \left( t \right)} \right\|\! \le \!\sum\limits_{q = 1}^i {\left\| {\Delta _q^\theta } \right\|h\left( {t - t_q^ + } \right)} \!\!  < \!\!\infty $. The requirements of Assumption 1 are not restrictive and usually satisfied in practical scenario.

\section{Main result}\label{sec3}

We propose to solve the problem \eqref{eq7} in two steps. At the first one, it is necessary to estimate the elements of the sequence \eqref{eq6}, i.e. to propose an \uline{estimation algorithm}. Using the obtained estimates $\hat t_i^+$, the second step is to derive an \uline{adaptive law} to track piecewise constant parameters of the regression \eqref{eq2} and ensure that the stated goal \eqref{eq7} is achieved. At the same time, the estimation algorithm and adaptive law should function in parallel in online mode and be robust to the possible presence of an external bounded disturbance in the regression \eqref{eq2}.

In Section 3.1 the estimation algorithm to detect time instants when the unknown parameters of regression \eqref{eq2} switch to new values is proposed, which ensures the required boundedness $\tilde t_i^ +  \le {T_i}$. In Section 3.2 an identification law is proposed. It is based on the obtained estimates $\hat t_i^ +$ and guarantees exponential convergence of $\tilde \theta \left( t \right)$ to zero under Assumption 1.

\subsection{Switching detection algorithm}\label{subsec31}

To introduce the algorithm of switching time instants detection, first of all, the dynamic regressor extension and mixing (DREM) procedure \cite{c17} is applied. In order to do that, the regression equation \eqref{eq5} is extended as:
\begin{eqnarray}\label{eq8}
\varphi \left( t \right)y\left( t \right) = \varphi \left( t \right){\varphi ^{\rm{T}}}\left( t \right){\theta \left( t \right)}{\rm{.}}
\end{eqnarray}

The filters with exponential forgetting and resetting at time point $\hat t_i^+$ are introduced:
\begin{eqnarray}\label{eq9}
\begin{array}{c}
\textcolor{black}{z}\left( t \right) = \int\limits_{\hat t_i^ + }^t {{e^{ - \int\limits_{\hat t_i^ + }^\tau  {\sigma ds} }}\varphi \left( \tau  \right)y\left( \tau  \right)} d\tau, \; z\left( {\hat t_i^ + } \right) = 0,\\
\omega \left( t \right) = \int\limits_{\hat t_i^ + }^t {{e^{ - \int\limits_{\hat t_i^ + }^\tau  {\sigma ds} }}\varphi \left( \tau  \right){\varphi ^{\rm{T}}}\left( \tau  \right)} {\rm{}}d\tau, \; \omega \left( {\hat t_i^ + } \right) = 0,
\end{array}
\end{eqnarray}
where $z\left( t \right) \in {\mathbb{R}^{n \times p}}{\rm{,}}\;\omega \left( t \right) \in {\mathbb{R}^{n \times n}}$. The time point $\hat t_i^+$ will be precisely defined further.

The dynamic regressor extension and mixing procedure \cite{c17} is applied to the function $z\left( t \right)$ to obtain:
\begin{eqnarray}\label{eq10}
\Upsilon \left( t \right){\rm{:}} = adj\left\{ {\omega \left( t \right)} \right\}z\left( t \right){\rm{,}}
\end{eqnarray}
where $\Upsilon \left( t \right) \in {\mathbb{R}^{n \times p}}$.

The expression $det \left\{ {\omega \left( t \right)} \right\}\theta \left( t \right)$  is added to and subtracted from \eqref{eq10}:
\begin{eqnarray}\label{eq11}
\begin{array}{c}
\Upsilon \left( t \right) = \Delta \left( t \right)\theta \left( t \right) + \varepsilon \left( t \right){\rm{,\;}}\\
\Delta \left( t \right) = det \left\{ {\omega \left( t \right)} \right\}, \; \varepsilon \left( t \right){\rm{:}} = adj\left\{ {\omega \left( t \right)} \right\}z\left( t \right) - \Delta \left( t \right)\theta \left( t \right){\rm{,}}
\end{array}
\end{eqnarray}
where $\Delta \left( t \right) \in \mathbb{R}$ is the measurable scalar regressor, $\varepsilon \left( t \right)$ is the unknown disturbance.

\textcolor{black}{The function $\varepsilon \left( t \right)$ is caused by fact that the commutative property is not applicable to the filtration \eqref{eq9} when $\theta \left( t \right)$ changes its value at the time instant $t_i^ +$, so $\varepsilon \left( t \right)$ is an indicator of the plant parameters change.}

\textcolor{black}{Then to estimate the time instants the function $\varepsilon \left( t \right)$ is to be expressed from \eqref{eq8} and \eqref{eq11}. To this end, the equation \eqref{eq8} is multiplied by the regressor $\Delta \left( t \right)$:
\begin{eqnarray}\label{eq12}
\Delta \left( t \right)\varphi \left( t \right)y\left( t \right) = \Delta \left( t \right)\varphi \left( t \right){\varphi ^{\rm{T}}}\left( t \right)\theta \left( t \right){\rm{,}}
\end{eqnarray}
whereas \eqref{eq11} – by $\varphi \left( t \right){\varphi ^{\rm{T}}}\left( t \right)$:
\begin{eqnarray}\label{eq13}
\varphi \left( t \right){\varphi ^{\rm{T}}}\left( t \right)\Upsilon \left( t \right) = \varphi \left( t \right){\varphi ^{\rm{T}}}\left( t \right)\Delta \left( t \right)\theta \left( t \right) + \varphi \left( t \right){\varphi ^{\rm{T}}}\varepsilon \left( t \right){\rm{,\;}}
\end{eqnarray}}

\textcolor{black}{The equation \eqref{eq12} is subtracted from \eqref{eq13} to obtain a measurable indicator of the regression equation \eqref{eq2} parameters change:
\begin{eqnarray}\label{eq14}
\epsilon \left( t \right){\rm{:}} = \varphi \left( t \right){\varphi ^{\rm{T}}}\left( t \right)\Upsilon \left( t \right) - \Delta \left( t \right)\varphi \left( t \right)y\left( t \right) = \varphi \left( t \right){\varphi ^{\rm{T}}}\left( t \right)\varepsilon \left( t \right){\rm{,\;}}\epsilon\left( {\hat t_i^ + } \right) = 0,
\end{eqnarray}
where $\epsilon \left( t \right) \in {\mathbb{R}^{n \times p}}$ is a residual.} 

So the following proposition is introduced on the basis of equation \eqref{eq14}.

\begin{proposition}\label{proC1}
If $\hat t_i^ +  \ge t_i^+$, then:
\begin{eqnarray}\label{eq15}
\epsilon \left( t \right){\rm{:}} = \left\{ \begin{array}{l}
\varphi \left( t \right){\varphi ^{\rm{T}}}\left( t \right)adj\left\{ {\omega \left( t \right)} \right\}\int\limits_{\hat t_{i - 1}^ + }^{t_i^ + } {{e^{ - \int\limits_{\hat t_{i - 1}^ + }^\tau  {\sigma ds} }}\varphi \left( \tau  \right){\varphi ^{\rm{T}}}\left( \tau  \right)} {\rm{}}\;d\tau \left( {{\theta _{i - 1}} - {\theta _i}} \right){\rm{,}}\;\forall t \in \left[ {t_i^ + {\rm{;}}\;\hat t_i^ + } \right)\\
{{0_{n \times p}}}{\rm{,}}\;\forall t \in \left[ {\hat t_i^ + {\rm{;}}\;t_{i + 1}^+} \right)
\end{array}. \right.
\end{eqnarray}

Proof of Proposition 1 is postponed to Appendix.
\end{proposition}

The results obtained in Proposition 1 make it reasonable to choose the function $\epsilon \left( t \right)$ as an indicator to find $\hat t_i^ +$  value. The following estimation algorithm is proposed to solve the problem under consideration:
\begin{eqnarray}\label{eq16}
\begin{array}{c}
{\rm{initialize:}}\; i \leftarrow 1,\; {t_{{\mathop{\rm up}\nolimits} }} = \hat t_{i - 1}^ + \\
{\rm{IF}}\; t - {t_{up}} \ge {\Delta _{pr}}\;{\rm{AND}}\; \epsilon\left\| {\left( t \right)} \right\| > 0\;\\
{\rm{THEN}}\;\hat t_i^ + {\rm{:}} = t + {\Delta _{pr}}, \; {t_{up}} \leftarrow t, \; i \leftarrow i + 1,
\end{array}
\end{eqnarray}
where $0 \le {\Delta _{pr}} < \mathop {\min }\limits_{\forall i \in \mathbb{N}} {T_i}$ is the parameter of the estimation algorithm.

Using the function \eqref{eq15}, the algorithm \eqref{eq16} detects the switching time instant with the precision ${\Delta _{pr}}$. The properties of the algorithm \eqref{eq16} are strictly described in the following proposition.

\begin{proposition}\label{proC2}
Let Assumption 1 be satisfied and $\hat t_i^+$ is estimated using \eqref{eq16}. Then $\hat t_i^ +  \ge t_i^ +$, and the condition $\tilde t_i^ +  \le {T_i}$ from \eqref{eq7} holds in case of the appropriate choice of  ${\Delta _{pr}}$.

Proof of Proposition 2 is presented in Appendix.
\end{proposition}

So, the estimation algorithm \eqref{eq16} detects the time points when the parameters switch their values to new ones in online mode. As a result, it allows one to obtain the estimation of the sequence \eqref{eq6}:
\begin{eqnarray}\label{eq17}
\hat \Im  = \left\{ {\hat t_0^ + {\rm{,}}\;\hat t_1^ + {\rm{}}\;{\rm{,}} \ldots {\rm{,}}\;\hat t_{i - 1}^ + {\rm{,}}\;\left. {\hat t_i^ + {\rm{,}} \ldots } \right|i \in \mathbb{N}} \right\}.
\end{eqnarray}

The estimation error between each element of \eqref{eq17} and \eqref{eq6} is equal or lower than the value of ${T_i}$ and could be adjusted by the appropriate choice of ${\Delta _{pr}}$. So, according to Proposition 2, it is concluded that, using the algorithm \eqref{eq16}, the first goal from \eqref{eq7} is achieved.

\emph{Remark 2:} It is of high importance for practice to ensure that the estimation algorithm \eqref{eq16} retains its properties in the presence of the external bounded disturbances in the regressions \eqref{eq2}, \eqref{eq5}, and \eqref{eq8}. It is easy to show that if $\forall t \ge t_0^ + {\rm{}}\;y\left( t \right) = {\varphi ^{\rm{T}}}\left( t \right){\theta \left( t \right)}{\rm{ + }}w\left( t \right){\rm{,}}$ then the residual \eqref{eq14} takes the form:
\begin{eqnarray}\label{eq18}
\begin{array}{l}
\epsilon \left( t \right){\rm{:}} = \varphi \left( t \right){\varphi ^{\rm{T}}}\left( t \right)adj\left\{ {\omega \left( t \right)} \right\}\int\limits_{\hat t_i^ + }^t {{e^{ - \int\limits_{\hat t_i^ + }^\tau  {\sigma ds} }}\varphi \left( \tau  \right){\varphi ^{\rm{T}}}\left( t \right)\theta \left( t \right)} {\rm{}}d\tau  - \Delta \left( t \right)\varphi \left( t \right){\varphi ^{\rm{T}}}\left( t \right)\theta \left( t \right) + \\
 + \varphi \left( t \right){\varphi ^{\rm{T}}}\left( t \right)adj\left\{ {\omega \left( t \right)} \right\}\int\limits_{\hat t_i^ + }^t {{e^{ - \int\limits_{\hat t_i^ + }^\tau  {\sigma ds} }}\varphi \left( \tau  \right)w\left( \tau  \right)} {\rm{}}d\tau  - \Delta \left( t \right)\varphi \left( t \right)w\left( t \right),\; \epsilon \left( {\hat t_i^ + } \right) = 0,
\end{array}
\end{eqnarray}
where $w\left( t \right) \in {\mathbb{R}^{m \times p}}$ is a bounded $\left\| {w\left( t \right)} \right\| \le {w_{\max }}$ external disturbance.

Hence, $\forall t \ne t_i^ + {\rm{\;}}\left\| {\epsilon \left( t \right)} \right\| > 0$. As a consequence, according to the algorithm \eqref{eq16}, the filters \eqref{eq9} reset their state periodically – each ${\Delta _{pr}}$ seconds. Such result is inappropriate from the practical point of view. So, using the mathematical statistics theory \cite{c47}, in case of external disturbances a modified (robust) version of the algorithm \eqref{eq16} should be used:
\begin{eqnarray}\label{eq19}
\begin{array}{c}
{\rm{initialize:}}\;i \leftarrow 1,{\rm{}}\;{t_{{\mathop{\rm up}\nolimits} }} = \hat t_{i - 1}^ + \\
{\rm{IF}}\;t - {t_{up}} \ge {\Delta _{pr}}{\rm{\;AND\;}}{\rm E}\left\{ {\epsilon \left( t \right)} \right\} > 0.9\sqrt {{\mathop{\rm var}} \left\{ {\epsilon \left( t \right)} \right\}}  + c\left( t \right){\rm{}}\\
{\rm{THEN}}\;\hat t_i^ + {\rm{:}} = t + {\Delta _{pr}}{\rm{,}}\;{t_{up}} \leftarrow t{\rm{,}}\;i \leftarrow i + 1,
\end{array}
\end{eqnarray}
where $c\left( t \right)$ is the arbitrary parameter of the robust algorithm.

The choice of the algorithm \eqref{eq19} parameter $c\left( t \right)$ value allows one to adjust the estimation accuracy and adapt to each specific class of the external disturbances. For example, if the disturbance is a noise with zero mean, then, according to the results \cite{c47, c48}, it is enough to choose $c = 0$. In the general case it is recommended to choose the function $c\left( t \right)$ as follows:
\begin{eqnarray}\label{eq20}
c\left( t \right) = E\left\{ {{w_{\max }}\varphi \left( t \right){\varphi ^{\rm{T}}}\left( t \right)adj\left\{ {\omega \left( t \right)} \right\}\int\limits_{\hat t_i^ + }^t {{e^{ - \int\limits_{\hat t_i^ + }^\tau  {\sigma ds} }}\varphi \left( \tau  \right)} {I_{m \times p}}{\rm{\;}}d\tau } \right\}
\end{eqnarray}

In the disturbance free case, the properties of the robust algorithm \eqref{eq19} completely coincide with the ones described in Proposition 2. However, if the external disturbance is added to the right-hand side of \eqref{eq2}, the algorithm \eqref{eq19}, in contrast to \eqref{eq16}, does not face problems of false detections in case the parameter $c\left( t \right)$ value is chosen correctly. It does not also cause periodic reset of the filter \eqref{eq9} and provides sufficient accuracy in terms of error $\tilde t_i^+$. More details about the robust algorithm \eqref{eq19} can be found in \cite{c47, c48}.

\emph{Remark 3:} The regression equation \eqref{eq2} is usually a parameterized form of representation of some particular adaptive control problems. It is often obtained using various stable minimum-phase filters (see \cite{c2} for details). Therefore, when the change of the regression \eqref{eq2} parameters is detected according to \eqref{eq16} or \eqref{eq19}, in addition to the filters \eqref{eq9}, all filters previously used to obtain the regression \eqref{eq2} must be set to their initial zero states.

\subsection{Adaptive law}\label{subsec32}

The required adaptive law of the unknown parameters $\theta \left( t \right)$ will be introduced on the basis of the regression function \eqref{eq10}, which is formed by the dynamic regressor extension and mixing procedure \cite{c17}. But before that let the properties of the function $\Upsilon \left( t \right)$ and the regressor $\Delta \left( t \right)$ be studied. The result of this analysis is presented in the form of a proposition.
\begin{proposition}\label{proC3}
Let the requirements of Assumption 1 be met, and $\hat t_i^ + $ be formed according to the algorithm \eqref{eq16}, then the regression function $\Upsilon \left( t \right)$ can be represented in the form:
\begin{eqnarray}\label{eq21}
\begin{array}{c}
\Upsilon \left( t \right){\rm{:}} = \Delta \left( t \right)\theta \left( t \right) + \varepsilon \left( t \right){\rm{,}}\\
\varepsilon \left( t \right){\rm{:}} = \left\{ \begin{array}{l}
adj\left\{ {\omega \left( t \right)} \right\}\int\limits_{\hat t_{i - 1}^ + }^{t_i^ + } {{e^{ - \int\limits_{\hat t_{i - 1}^ + }^\tau  {\sigma ds} }}\varphi \left( \tau  \right){\varphi ^{\rm{T}}}\left( \tau  \right)}\;d\tau \left( {{\theta _{i - 1}} - {\theta _i}} \right), \; \forall t \in \left[ {t_i^ + {\rm{;\;}}\hat t_i^ + } \right){\rm{,}}\\
0_{n \times p},\;\forall t \in \left[ {\hat t_i^ + {\rm{;\;}}t_{i + 1}^ + } \right){\rm{,}}
\end{array} \right.
\end{array}
\end{eqnarray}
where $\left\| {\varepsilon\left( t \right)} \right\| \le {\varepsilon_{max}}$, and the regressor $\Delta \left( t \right)$ is such that $\forall t \in \left[ {t_i^ + {\rm{ + }}{T_i}{\rm{;}}\;\hat t_{i + 1}^ + } \right)\;{\Delta _{UB}} \ge \Delta \left( t \right) \ge {\Delta _{LB}} > 0$.

Proof of Proposition 3 and the definitions of ${\varepsilon_{max}},\;{\Delta _{LB}}, \; {\Delta _{UB}}$ are presented in Appendix.
\end{proposition}

According to the above-stated analysis, considering the conservative case, the regressor $\Delta \left( t \right)$ may become singular $\Delta \left( t \right) = 0$ over the time range $\left[ {t_i^ + \; t_i^ + {\rm{ + }}{T_i}} \right)$ and, as a consequence, it does not meet the necessary condition of the exponential stability \eqref{eq2}.

Because of that, to bound the regressor $\Delta \left( t \right)$ away from zero globally from the time instant $t = t_0^ + {\rm{ + }}{T_0}$, the following filtration of the equation \eqref{eq21} is introduced into consideration
\begin{subequations}
\begin{align} \label{eq22a}
\dot {\cal Y}\left( t \right) =  - k\left( {{\cal Y}\left( t \right) - \Upsilon \left( t \right)} \right), \; {\cal Y}\left( {t_0^ + } \right) = {0_{n \times p}}{\rm{,}}
\end{align}
\begin{align} \label{eq22b} 
\dot \Omega \left( t \right) =  - k\left( t \right)\left( {\Omega \left( t \right) - \Delta \left( t \right)} \right), \; \Omega \left( {t_0^ + } \right) = 0,
\end{align}
\begin{align} \label{eq22c}
{\dot \varepsilon _f}\left( t \right) =  - k\left( {{\varepsilon _f}\left( t \right) - \varepsilon \left( t \right)} \right), \;{\varepsilon _f}\left( {t_0^ + } \right) = {0_{n \times p}}{\rm{,}}
\end{align}
\end{subequations}
where $ k > 0$.

Then, due to the fact that the commutative property is not applicable to the filter (22a), by analogy with \eqref{eq11}, it is obtained: 
\begin{eqnarray}\label{eq23}
\begin{array}{c}
{\cal Y}\left( t \right) = \Omega \left( t \right)\theta \left( t \right) + d\left( t \right){\rm{,}}\\
d\left( t \right) = {\cal Y}\left( t \right) - \Omega \left( t \right)\theta \left( t \right), \; d\left( {t_0^ + } \right) = {0_{n \times p}}.
\end{array}
\end{eqnarray}
The following proposition holds for the new regressor $\Omega \left( t \right)$.
\begin{proposition}\label{proC4}
Let the requirements of Assumption 1 be met, and $\hat t_i^ +$  is obtained according to \eqref{eq16}, then:
\begin{center}$\forall t \ge t_0^ + {\rm{ + }}{T_0}\;\;{\Omega _{UB}} \ge \Omega \left( t \right) \ge {\Omega _{LB}} > 0.$\end{center}

Proof of Proposition 4 is presented in Appendix.
\end{proposition}

Taking into account that ${\cal Y}\left( t \right)$ can be represented as \eqref{eq23}, as well as the properties $\Omega \left( t \right)$, the unknown parameters adaptive law is chosen as:
\begin{eqnarray}\label{eq24}
\begin{array}{c}
\dot {\hat \theta} \left( t \right){\rm{:}} =  - \gamma \Omega \left( t \right)\left( {\Omega \left( t \right)\hat \theta \left( t \right) - {\cal Y}\left( t \right)} \right) =  - \gamma {\Omega ^2}\left( t \right)\tilde \theta \left( t \right) + \gamma \Omega \left( t \right)\varepsilon\left( t \right){\rm{,}}\\
\gamma  = \left\{ \begin{array}{l}
0,\;if \; \Omega \left( t \right) \le \rho {\rm{,}}\\
{\textstyle{{{\gamma _0}} \over {{\Omega ^2}\left( t \right)}}}, \; otherwise.
\end{array} \right.
\end{array}
\end{eqnarray}
where $0 < {\gamma _0} \le k$  is the adaptive gain, $\rho  \in \left( {0;\; {\Omega _{LB}}} \right]$ is the arbitrary parameter of the adaptive law.

The properties of the error $\tilde \theta \left( t \right)$ are analyzed in Theorem.
\begin{theorem}\label{thm1}
Let the requirements of Assumption 1 be satisfied, $\hat t_i^ +$ be obtained in accordance with \eqref{eq16} under ${\Delta _{pr}} \to 0$, and $\hat \theta \left( t \right)$ be calculated using \eqref{eq24}, then $\forall t \ge t_0^ + {\rm{ + }}{T_0}$ the parameter error $\tilde \theta \left( t \right)$ converges exponentially to zero with the rate that is proportional to ${\gamma _0}$.

The proof of the Theorem is given in Appendix.
\end{theorem}
Thus, if Assumption 1 requirements are met, the adaptive law \eqref{eq24} provides exponential convergence of the parameters error to zero with the rate adjusted with the help of $ {\gamma _0}$. The proposed algorithm for piecewise constant parameters identification (see Algorithm 1) consists of the procedure of dynamic regressor extension and mixing \eqref{eq9}, \eqref{eq10}, residual calculation equation \eqref{eq14}, algorithm to detect parameters switches \eqref{eq16} or \eqref{eq19}, filtering \eqref{eq22a}, \eqref{eq22b} and adaptive law \eqref{eq24}. \textcolor{black}{The block diagram of the developed system is shown in Fig. 1.}
\begin{figure}[h]
\centerline{\includegraphics[height=7pc]{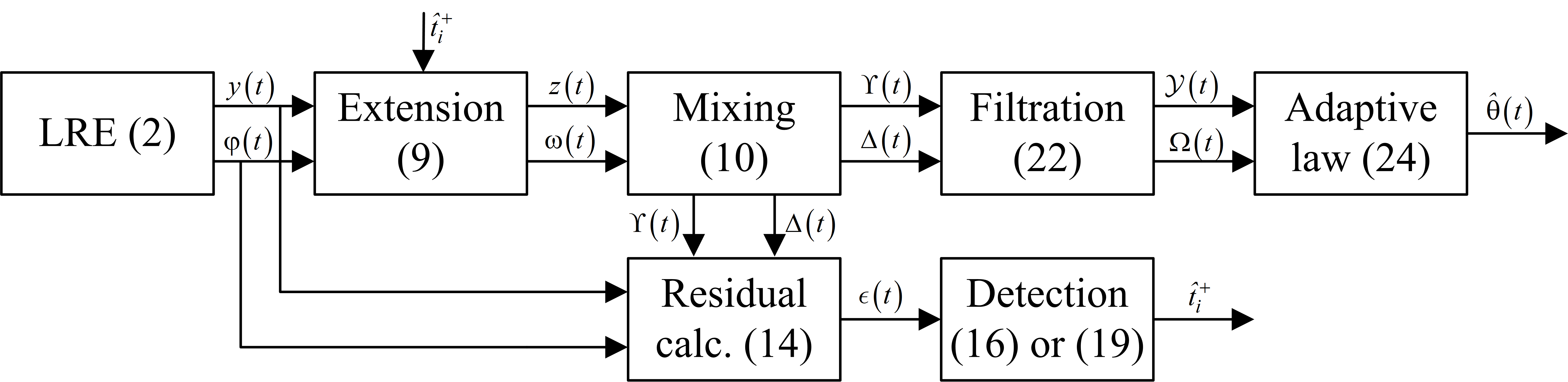}}
\caption{\textcolor{black}{Block diagram of the proposed identifier}\label{fig1}}
\end{figure}

\begin{algorithm}
\caption{\textcolor{black}{Pseudocode of algorithm to identify PC parameters}}\label{alg1}
\begin{algorithmic}[1]

\State \textbf{Input:}  $i,\; \hat{t}_{up},\;t_{i}^+,\;\varphi(t),\;y(t),\;k,\;\rho, \;\sigma,\;\gamma_0,\; w_{max}$
\State \textbf{Output:}  $i,\; t_{up},\;\hat{t}_{i}^+, \hat{\theta}(t)$
\State $z\left( t \right),\; \omega \left( t \right) \Leftarrow eq.{{\eqref{eq8}\;and\;eq.\eqref{eq9}}}$
\State $\Upsilon \left( t \right),\;\Delta \left( t \right) \Leftarrow eq.{\eqref{eq10}}\;and\;eq.{\eqref{eq11}}$
\State $\epsilon \left( t \right) \Leftarrow eq.{{\eqref{eq14}}}$
   \If{${w_{\max }} \ne 0$}
   \State ${E}\left\{\epsilon{\left( t \right)} \right\},\; {\mathop{var}} \left\{\epsilon{\left( t \right)} \right\}$
  \State {$c\left( t \right) \Leftarrow eq.\eqref{eq20}$} 
   \EndIf
\If{$\left( {{w_{\max }} = 0\;{\rm{\textbf{and}}}\;t - {t_{up}} \ge {\Delta _{pr}}\;{\rm{\textbf{and}}}\;\left\|\epsilon{\left( t \right)} \right\| > 0} \right)\;{\rm{\textbf{or}}}\;\left( {{w_{\max }} \ne 0{\rm{\;\textbf{and}\;}}t - {t_{up}} \ge {\Delta _{pr}}{\rm{\;\textbf{and}\;}}{\rm E}\left\{\epsilon{\left( t \right)} \right\} > 0.9\sqrt {{\mathop{var}} \left\{\epsilon{\left( t \right)} \right\}}  + c\left( t \right)} \right)$}
 $\hat t_i^ + {\rm{:}} = t + {\Delta _{pr}}{\rm{,\;}}{t_{up}} \leftarrow t{\rm{,\;}}i \leftarrow i + 1$ 
 \EndIf
 \State $\Omega \left( t \right),{\rm{\;}}{\cal Y}\left( t \right) \Leftarrow eq.{\eqref{eq22a}}\; and \; eq.{\eqref{eq22b}}$
 \State $\dot {\hat \theta} \left( t \right) \Leftarrow eq.{\eqref{eq24}}$
 \State $\textbf{Return:}\;i,\; {t_{{\mathop{\rm up}\nolimits}}},\;\hat t_i^ +,\;\hat\theta \left( t \right)$
\end{algorithmic}
\end{algorithm}

\textcolor{black}{Estimating the computational complexity of the proposed algorithm, it should be noted that its upper bound is defined by the complexity of the most computationally intensive operations. They include the calculation of the $adj\left\{ {\omega \left( t \right)} \right\}$ and $det \left\{ {\omega \left( t \right)} \right\}$, as well as matrix multiplication. Considering the first two operations, their complexity is $O\left( {{n^3}} \right)$, for square matrices multiplication -- $O\left( {{n^3}} \right)$. For non-square matrices such estimation is $O\left( {n \cdot m \cdot p} \right)$, but without losing generality, we can assume that $n$ has the greatest value among $n, m$ and $p$, then the estimation $O\left( {n \cdot m \cdot p} \right) < O\left( {{n^3}} \right)$ can be used for non-square matrices as well. Thus, the upper bound on the computational complexity of the proposed identification algorithm is $O\left( {{n^3}} \right)$.}

\emph{Remark 4:} The operation of division by ${\Omega ^2}\left( t \right)$ used in \eqref{eq24} is “safe” one as, if Assumption 1 requirements are met, then, in accordance with Proposition 4, $\forall t \ge t_0^ + {\rm{ + }}{T_0}{\rm{\;}}\Omega \left( t \right) \ge {\Omega _{LB}} > 0$. Also, only one switch of the nonlinear operator \eqref{eq24} happens $\forall t \ge t_0^ +  + {T_0}$, particularly, at the time point $t_0^ + {\rm{ + }}{T_0}$. Such division by ${\Omega ^2}\left( t \right)$ allows one to provide the desired rate (defined by ${\gamma _0}$) of the exponential convergence of the identification error $\tilde \theta \left( t \right)$. However, to use the division the value of the parameter $\rho$ is required to be chosen in accordance with $\rho  \in \left( {0{\rm{;\;}}{\Omega _{LB}}} \right)$.

Actually, the adaptive law \eqref{eq24} can be rewritten in the following well-known \cite{c50} form:
\begin{eqnarray}\label{eq25}
\begin{array}{c}
{{\hat \theta }_{FT}}\left( t \right){\rm{:}} = \left\{ \begin{array}{l}
\hat \theta \left( t \right){\rm{,}}\;if{\rm{}}\;\Omega \left( t \right) \le \rho \\
\frac{{{{\cal Y}} \left( t \right)}}{{\Omega \left( t \right)}}{\rm{,}}\;otherwise
\end{array}, \right.\\
\dot {\hat \theta} \left( t \right) =  - {\gamma _0}\hat \theta \left( t \right) + {\gamma _0}{{\hat \theta }_{FT}}\left( t \right){\rm{,}}
\end{array}
\end{eqnarray}
where ${\hat \theta _{FT}}\left( t \right)$ is the finite time estimation of the unknown parameters, and $\gamma_0$  has the sense of an aperiodic filter constant.

\emph{Remark 5:} Let a bounded external disturbance be added to \eqref{eq2}, so that $\forall t \ge t_0^ +$ \linebreak $y\left( t \right) = {\varphi ^{\rm{T}}}\left( t \right){\theta \left( t \right)}{\rm{ + }}w\left( t \right)$. Then, in accordance with the results of \cite{c51} and due to boundedness of $w\left( t \right)$ and $\varphi \left( t \right)$, the filtrations \eqref{eq9} and \eqref{eq22a}-\eqref{eq22c} form a function ${\cal Y}\left( t \right)$ with the bounded external disturbance  $W\left( t \right)$:
\begin{eqnarray}\label{eq26}
{\cal Y}\left( t \right){\rm{:}} = \Omega \left( t \right)\theta \left( t \right) + d\left( t \right) + {\mathop{W}\nolimits} \left( t \right){\rm{.}}
\end{eqnarray}

In such a case, the adaptive law \eqref{eq24} is rewritten as:
\begin{eqnarray}\label{eq27}
\begin{array}{c}
\dot {\hat \theta} \left( t \right) =  - \gamma {\Omega ^2}\left( t \right)\hat \theta \left( t \right) + \gamma {\Omega ^2}\left( t \right)\theta \left( t \right) + \gamma \Omega \left( t \right)\left( {d\left( t \right) + W\left( t \right)} \right){\rm{,}}\\
\gamma  = \left\{ \begin{array}{l}
0,{\rm{}}\;if{\rm{}}\;\Omega \left( t \right) \le \rho\\
{\textstyle{{{\gamma _0}} \over {{\Omega ^2}\left( t \right)}}}{\rm{,}}\;otherwise
\end{array}, \right.
\end{array}
\end{eqnarray}
and, because $W\left( t \right)$ is not necessarily exponentially vanishing and bounded only, instead of the exponential convergence of $\tilde \theta \left( t \right)$ to zero, it provides the exponential convergence of $\tilde \theta \left( t \right)$ to a compact set: $\left\| {\tilde \theta \left( t \right)} \right\| \le \left\| {W\left( t \right)} \right\|{\Omega ^{ - 1}}\left( t \right).$

\section{Discussion}\label{sec4}
\textcolor{black}{In this section, firstly, the guideline to choose the values of the proposed identification procedure parameters is described, and secondly, the adaptive law \eqref{eq24} is compared with some of the known ones in the literature.}
\subsection{Tuning guidelines}\label{sec41}\textcolor{black}{
Arbitrary parameters of the proposed identification algorithm can be divided into three main groups:}
\begin{itemize}
\item[$(i)$] \textcolor{black}{parameters $\sigma, \; k$ to provide the propagation of the regressor excitation in parameterizations \eqref{eq9}, \eqref{eq22a}-\eqref{eq22c}};

\item[$(ii)$] \textcolor{black}{detection algorithm \eqref{eq16} or \eqref{eq19} parameters ${\Delta _{pr}, \; {w_{\max }}}$;}
\item[$(iii)$]\textcolor{black}{the parameters ${\gamma _0},\;\rho $ of the adaptive law \eqref{eq24}.}
\end{itemize}

\textcolor{black}{The basic rules and prerequisites to choose the values of the parameters are considered separately for each category.}

\subsubsection{Guide to choose $\sigma$ and $k$}\label{sec411}

\textcolor{black}{The parameter $\sigma$ defines the length of the time interval when \eqref{eq9} is sensitive to new input data over the range $\left[ {\hat t_i^ + {\rm{; \;}}\hat t_{i + 1}^ + } \right]$. When the value of $\sigma$  is low, the filter \eqref{eq9} is an open-loop integrator, which is sensitive to new input data almost all the time, whereas, when $\sigma$ is high, it is an integrator with an input damping and keeps alertness only over some time interval. By varying the width of the sensitivity time window with the help of $\sigma$ value, it is possible to significantly influence the quality of transients of the unknown parameters estimates $\hat \theta \left( t \right)$. For example, if $\sigma$ is chosen such that the time instant $t_i^ +$ falls into the region of insensitivity of \eqref{eq9}, then over the interval $\left[ {t_{i + 1}^ + {\rm{; \;}}\hat t_{i + 1}^ + } \right]$ the law \eqref{eq24} identifies the parameters ${\theta _i}$ instead of their actual value ${\theta _{i + 1}}$. If, on the other hand, the time instant $t_i^ +$ falls into the region of sensitivity, a superpositional mixing of data on the current ${\theta _{i + 1}}$ and previous ${\theta _i}$ values of the regression parameters occurs over the interval $\left[ {t_{i + 1}^ + {\rm{; \;}}\hat t_{i + 1}^ + } \right]$, and the transient of $\hat \theta \left( t \right)$ can contain both peaking phenomena and oscillations. In the case when an external bounded disturbance $w\left( t \right)$ affects the regression \eqref{eq2}, the choice of $\sigma$ also allows one to adjust the ratio “disturbance/regressor -- $\left\| {W\left( t \right)} \right\|{\Omega ^{ - 1}}\left( t \right)$” in equation \eqref{eq23} and, consequently, influence the quality of $\hat \theta \left( t \right)$ transients.}

\textcolor{black}{Thus, summarizing the above-stated facts, it is recommended to choose the value of the parameter $\sigma$ in accordance with the equation $\sigma  = \left( {3 \div 5} \right)T_{\min }^{ - 1}$, using {\it{a priori}} assumptions on the minimum possible time ${T_{\min }}$ between two consecutive switches.}

\textcolor{black}{The parameter $k$ is a coefficient of a conventional first order filter and defines the rate of convergence of ${\cal Y}\left( t \right)$ and $\Omega \left( t \right)$ to $\Upsilon \left( t \right){\rm{,\;}}\Delta \left( t \right)$ respectively. The parameter $k$ sets the lower bound on the rate of $\tilde \theta \left( t \right)$ exponential convergence to zero, and therefore must be chosen on the basis of the desired performance of the piecewise constant parameters identification procedure and the required rate of convergence to zero. It is also important to note that, considering the case of an infinite number of the regression \eqref{eq2} parameters switches, the parameter $k$ defines a majorant function for the regression parameters correction (see statement 5.2 of Assumption 1).}

\subsubsection{Guide to choose ${\Delta _{pr}}$ and ${w_{\max}}$}\label{sec412}

\textcolor{black}{The parameter ${\Delta _{pr}}$, first of all, defines a small neighborhood of the time point $t_i^ +$, in which no change of the regression \eqref{eq2} parameters can happen $\left( {t_i^ +  + {\Delta _{pr}} \ll t_i^ +  + {T_{\min }}} \right)$. Secondly, ${\Delta _{pr}}$ separates the time point of detection \eqref{eq16} and the one to reset the filters \eqref{eq9} in order to make the computational procedures more stable. In practice, even when ${\Delta _{pr}} = 0$, there is some time delay between the detection of the parameters change and the filters \eqref{eq9} resetting. Therefore, as far as the mathematical analysis is concerned, it is considered that the parameter ${\Delta _{pr}}$ allows one to take into account, in some way, what effect such delay has on the quality of the estimates. However, it should be noted that the above-stated reasoning is also valid when ${\Delta _{pr}} = 0$, and in such case the algorithm \eqref{eq16} provides the best detection accuracy $\hat t_i^ +  \to t_i^ +$. Thus, the parameter ${\Delta _{pr}}$ should be chosen on the basis of the desired performance of the detection algorithm of the unknown parameters switches.}

\textcolor{black}{The parameter ${w_{\max }}$ defines the robust properties of the detection algorithm \eqref{eq19} and should be chosen on the basis of {\it{a priori}} assumptions on $w\left( t \right)$.}

\subsubsection{Guide to choose ${\gamma _0}$ and $\rho$}\label{sec413}
\textcolor{black}{In accordance with the proof of Theorem, the parameter ${\gamma _0}$ defines the exponential convergence rate of the parameter error to zero, and therefore, as well as the parameter $k$, is to be chosen on the basis of the desired performance of the identification system and the required time of $\tilde \theta \left( t \right)$ convergence to a neighborhood of zero.}

\textcolor{black}{The parameter $\rho$, first of all, starts the procedure of the regressor excitation normalization \eqref{eq24}, which in turn allows one to ensure exponential convergence of $\tilde \theta \left( t \right)$ to zero with the rate, which is defined by the parameter ${\gamma _0}$, regardless of the $\Omega \left( t \right)$, and secondly, enables one to use the adaptive law \eqref{eq24} only when the regressor is sufficiently exciting. When there are no external disturbances, the value of the parameter $\rho$ defines the time instant when the identification process of the unknown parameters starts. In this case, in order to ensure exponential stability, it is necessary to choose $\rho$ such that $\rho  \in \left( {0{\rm{;\;}}{\Omega _{LB}}} \right]$. This can be done using the {\it{a priori}} conservative estimates of the regressor $\Omega \left( t \right)$ value. Given that the external disturbances affect the regression, the correct choice of the parameter $\rho$ allows one to start the process of unknown parameters identification only when the ratio $\left\| {W\left( t \right)} \right\|{\Omega ^{ - 1}}\left( t \right)$ has become sufficiently close to zero. Owing to this effect, it is possible to significantly improve the quality of $\hat \theta \left( t \right)$ transients in case of disturbances.}

\subsection{Comparison with Other Approaches}\label{sec41}
\textcolor{black}{The adaptive law \eqref{eq24} is compared with some identification methods of piecewise constant unknown parameters of the linear regression equation. Most of the composite adaptation methods \cite{c6, c7, c8, c9, c10, c11, c12, c13, c14, c15, c16, c17, c18, c19, c20, c21, c22} are intentionally excluded from the consideration since a detailed discussion and experimental demonstration of their shortcomings are given in \cite{c33}. Additionally, the offline or discrete identification methods \cite{c38, c39, c40}are also not considered as the main motivation of the piecewise constant parameter estimation laws development is their further application to online adaptive control and continuous-time observation problems.}
\subsubsection{Gradient Descent Law for PC Parameters Identification}\label{sec421}

\textcolor{black}{According to the gradient-based method of the unknown piecewise constant parameters identification, $N$ adjustable models are introduced \cite{c41}:
\begin{eqnarray}\label{eq28}
{\hat y_j}\left( t \right) = \left\{ \begin{array}{l}
{\varphi ^{\rm{T}}}\left( t \right){{\hat \Theta }_j}\left( t \right){\rm{,\;if\;}}\kappa \left( t \right) = j,\\
y\left( t \right){\rm{,\;if\;}}\kappa \left( t \right) \ne j.
\end{array} \right.
\end{eqnarray}}

\textcolor{black}{Considering \eqref{eq28} and \eqref{eq2}, the following law is used to obtain the estimate ${\hat \Theta _j}\left( t \right)$:
\begin{eqnarray}\label{eq29}
{\dot {\hat {\Theta}} _j}\left( t \right) = \left\{ \begin{array}{l}
 - \Gamma \varphi \left( t \right)\left( {{\varphi ^{\rm{T}}}\left( t \right){{\hat \Theta }_j}\left( t \right) - y\left( t \right)} \right){\rm{,\;if\;}}\kappa \left( t \right) = j,\\
0,{\rm{\;if\;}}\kappa \left( t \right) \ne j
\end{array} \right.
\end{eqnarray}
where $\Gamma  \in {\mathbb{R}^{n \times n}}$ is the adaptive gain matrix.}

\textcolor{black}{The law \eqref{eq24} to trace the piecewise constant function $\theta \left( t \right) = {\Theta _{\kappa \left( t \right)}}$ uses a single general vector of continuous estimate $\hat \theta \left( t \right)$, whereas \eqref{eq29} exploits the piecewise-continuous function ${\hat \Theta _{\kappa \left( t \right)}}$ obtained from the continuous estimates ${\hat \Theta _j}\left( t \right)$ of the parameters of each model \eqref{eq28}. This requires {\it{a priori}} knowledge of time instants $t_i^ +$ and function $\kappa \left( t \right)$ and represents a fundamental difference between \eqref{eq24} and \eqref{eq29}.}

\textcolor{black}{If the requirement of persistent excitation for $\varphi \left( t \right)$ is met, and each of the models \eqref{eq28} is active for sufficiently long period of time (statement 1 of Assumption 1), then \eqref{eq29} guarantees exponential convergence of the error ${\tilde \Theta _j}\left( t \right)$ to zero in case of any switching sequence $\kappa \left( t \right)$. In turn, the convergence of ${\tilde \Theta _j}\left( t \right)$ results in exponential convergence of the piecewise-continuous estimate ${\hat \Theta _{\kappa \left( t \right)}}\left( t \right)$ to the function ${\Theta _{\kappa \left( t \right)}}$.}

\textcolor{black}{As proved in Theorem, if $\varphi \left( t \right) \in \rm{FE}$ after each change of the regression parameters (statement 2 of Assumption 1), the law \eqref{eq24} ensures a global exponential convergence of the error $\tilde \theta \left( t \right) = \hat \theta \left( t \right) - \theta \left( t \right) = \hat \theta \left( t \right) - {\Theta _{\kappa \left( t \right)}}$ to zero. The finite excitation condition requires the regressor to be exciting after each regression parameters change over some time range only, while the persistent excitation requirement demands the regressor to be exciting over the whole regular time interval between the consecutive parameters changes. Therefore, the convergence condition of the law \eqref{eq24} is strictly weaker than the one of the \eqref{eq29}.}

\textcolor{black}{Thus, the differences between \eqref{eq24} and \eqref{eq29} are summarized as follows:}
\begin{itemize}
\item[--] \textcolor{black}{the knowledge of $t_i^ +$ and $\kappa \left( t \right)$ is not required to implement \eqref{eq24};}
\item[--] \textcolor{black}{the law \eqref{eq29} requires $\forall j \in \overline {1,N}$ to store ${\hat \Theta _j}\left( t \right)$, but this is not reasonable in case $N \to \infty$;}
\item[--] \textcolor{black}{the law \eqref{eq24} guarantees the exponential stability under weaker excitation condition;}
\item[--]  \textcolor{black}{the vector of estimates ${\hat \Theta _{\kappa \left( t \right)}}\left( t \right)$ of the piecewise constant parameters ${\Theta _{\kappa \left( t \right)}}$ is a discontinuous function;}
\item[--] \textcolor{black}{the vector of estimates $\hat \theta \left( t \right)$ of the piecewise constant parameters $\theta \left( t \right) = {\Theta _{\kappa \left( t \right)}}$ is a continuous function;}
\item[--] \textcolor{black}{the law \eqref{eq29} is insensitive to parameters changes in case they are not determined by the known function $\kappa \left( t \right)$.}
\end{itemize}

\subsubsection{Concurrent learning for PC Parameters Identification}\label{sec422}

\textcolor{black}{The following concurrent-learning-based law \cite{c35, c41, c44} allows one to overcome the main disadvantage of the gradient-based law \eqref{eq29} and ensure the exponential convergence of ${\tilde \Theta _j}\left( t \right)$ without the PE condition:
\begin{eqnarray}\label{eq30}
{\dot {\hat {\Theta }}_j}\left( t \right) = \left\{ \begin{array}{l}
 - {\Gamma _1}\varphi \left( t \right)\left( {{\varphi ^{\rm{T}}}\left( t \right){{\hat \Theta }_j}\left( t \right) - y\left( t \right)} \right) - {\Gamma _2}\sum\limits_{k = 1}^\ell  {{\varphi _k}\left( {\varphi _k^{\rm{T}}{{\hat \Theta }_j}\left( t \right) - {y_k}} \right)}  = \\
 =  - {\Gamma _1}\varphi \left( t \right)\left( {{\varphi ^{\rm{T}}}\left( t \right){{\hat \Theta }_j}\left( t \right) - y\left( t \right)} \right) - {\Gamma _2}{\cal R}{{\tilde \Theta }_j}\left( t \right){\rm{,\;if\;}}\kappa \left( t \right) = j\\
 - {\Gamma _2}{\cal R}{{\tilde \Theta }_j}\left( t \right){\rm{,\;if\;}}\kappa \left( t \right) \ne j
\end{array} \right.,
\end{eqnarray}
where ${y_k} = y\left( {{t_k}} \right){\rm{,\;}}{\varphi _k} = \varphi \left( {{t_k}} \right)$ are the values of the function $y\left( t \right)$ and the regressor $\varphi \left( t \right)$ respectively at the time instant ${t_k}$, which meets the condition $\forall k \in \overline {1,\ell } {\rm{,\;}}{t_k} \in \left[ {t_{{j_i}}^ + {\rm{;\;}}t_{{j_i} + 1}^ + } \right)$, ${\cal R} = \sum\limits_{k = 1}^\ell  {{\varphi _k}\varphi _k^{\rm{T}}}  \ge 0$ is a full-rank matrix as the regressor $\varphi \left( t \right)$ is finitely exciting after each activation of the ${j^{{\rm{th}}}}$ model.}

\textcolor{black}{To define the time instants ${t_k}$, at which it is reasonable to save the values of $y\left( {{t_k}} \right)$ and $\varphi \left( {{t_k}} \right)$ into the data stack, the algorithm of minimum eigenvalue ${\cal R}$ maximization is usually used \cite{c6}.}

\textcolor{black}{In contrast to \eqref{eq29}, the combined law \eqref{eq30} guarantees exponential convergence of all errors ${\tilde \Theta _j}\left( t \right)$ to zero if each ${j^{{\rm{th}}}}$ model has been active at least once, and after its activation the regressor finite excitation condition was met over the time range $\left[ {t_{{j_i}}^ + {\rm{;\;}}t_{{j_i} + 1}^ + } \right)$ of ${\cal R}$ calculation. Consequently, the main difference between \eqref{eq30} and \eqref{eq29} is the adjustment of the ${j^{{\rm{th}}}}$ model parameters even after its deactivation.}

\textcolor{black}{The convergence conditions of \eqref{eq24} and \eqref{eq30} coincide, and \eqref{eq30} guarantees exponential convergence of the piecewise-continuous estimate ${\hat \Theta _{\kappa \left( t \right)}}$ to the piecewise constant function ${\Theta _{\kappa \left( t \right)}}$ when the regressor is FE after each model activation.}

\textcolor{black}{Thus, the differences between \eqref{eq24} and \eqref{eq30} are as follows:}
\begin{itemize}
\item[--] \textcolor{black}{\eqref{eq24} obtains the continuous estimates $\hat \theta \left( t \right)$ to track the piecewise constant function $\theta \left( t \right) = {\Theta _{\kappa \left( t \right)}}$;}
\item[--]  \textcolor{black}{\eqref{eq30} requires $t_i^ +$ and $\kappa \left( t \right)$ to be known;}
\item[--] \textcolor{black}{\eqref{eq24} uses only current data on $y\left( t \right)$ and $\varphi \left( t \right)$;}
\item[--] \textcolor{black}{\eqref{eq30} requires off-line updates of the estimates ${\hat \Theta _j}\left( t \right)$, which is not reasonable in case $N \to \infty$;}
\item[--] \textcolor{black}{\eqref{eq30} is insensitive to parameters changes in case they are not determined by the known function $\kappa \left( t \right)$.}
\end{itemize}

\textcolor{black}{Here it is necessary to note the existence of improved concurrent-learning-based adaptation laws \cite{c45, c52}, in which, together with the calculation of ${\hat \Theta _j}\left( t \right)$, the time instants $t_i^ +$ are detected and the current state $\kappa \left( t \right)$ is identified. The disadvantages of such improved methods are the necessity to store and process off-line large amount of data on $y\left( t \right)$, $\varphi \left( t \right)$ and ${\hat \Theta _j}\left( t \right)$ for different values of $\kappa \left( t \right)$.}

\textcolor{black}{In general, the main advantage of the proposed law \eqref{eq24} over \eqref{eq29} and \eqref{eq30} is that it does not use memory stacks. The law \eqref{eq24} functions online and does not store/process offline any data on function \eqref{eq2}, whereas it is a must for \eqref{eq29} and \eqref{eq30} to store at least information about the parameters ${\hat \Theta _j}\left( t \right)$ of inactive models \eqref{eq28}. However, this advantage, which is important from the point of view of practical implementation, is obtained at the expense of additional conditions (see statement 5 from Assumption 1) being imposed on the number of switches of regression unknown parameters or the amplitude of the unknown parameters change.}

\textcolor{black}{Then the law \eqref{eq24} is compared with methods of piecewise constant parameters identification, which, as well as \eqref{eq24}: a) do not introduce $N$ adjustable models into consideration and use only one law to obtain continuous estimates $\hat \theta \left( t \right)$ to track $\theta \left( t \right) = {\Theta _{\kappa \left( t \right)}}$;  b) do not require knowledge about $t_i^ +$ and $\kappa \left( t \right)$.}

\subsubsection{Concurrent learning with Data Stack Purging}\label{sec423}

\textcolor{black}{One of the first laws proposed for continuous identification of the piecewise constant unknown parameters ${\Theta _{\kappa \left( t \right)}}$ when $t_i^ +$ and $\kappa \left( t \right)$ are unknown is a concurrent-learning-based law with stack purging \cite{c41, c53}:
\begin{eqnarray}\label{eq31}
\dot {\hat{\theta}} \left( t \right) =  - {\Gamma _1}\varphi \left( t \right)\left( {{\varphi ^{\rm{T}}}\left( t \right)\hat \theta \left( t \right) - y\left( t \right)} \right) - {\Gamma _1}\sum\limits_{k = 1}^\ell  {{\varphi _k}\left( {\varphi _k^{\rm{T}}\hat \theta \left( t \right) - {y_k}} \right)}  =  - {\Gamma _1}\varphi \left( t \right)\left( {{\varphi ^{\rm{T}}}\left( t \right){{\hat \Theta }_j}\left( t \right) - y\left( t \right)} \right) - {\Gamma _1}{\cal R}\tilde \theta \left( t \right),
\end{eqnarray}
where ${\cal R}$ and ${\varphi _k}{\rm{,\;}}{y_k}$ are updated with the help of the following purging algorithm:
\begin{eqnarray}\label{eq32}
{\lambda _{\min }}\left( {{{\cal R}_{new}}} \right) > {c_1}{\lambda _{\min }}\left( {{{\cal R}_{use}}} \right){e^{ - {c_2}\left( {t - \bar t} \right)}} + {c_3} \Rightarrow {{\cal R}_{use}} = {{\cal R}_{new}}{\rm{,\;}}\varphi _k^{use} = \varphi _k^{new}{\rm{,\;}}y_k^{use} = y_k^{old}{\rm{,}}
\end{eqnarray}
where ${c_1}{\rm{,\;}}{c_2}{\rm{,\;}}{c_3}$ are some arbitrary constants, $\bar t$ is the time instant of previous purging.}

\textcolor{black}{Purging \eqref{eq32} allows one to get rid of the outdated data accumulated in the stack and thereby ensure the sensitivity of the law \eqref{eq31} to changes of the values of the regression \eqref{eq2} parameters. However, in \cite{c41, c53}, when the stack purging algorithm \eqref{eq32} is used, the exponential convergence of the error $ \tilde \theta \left( t \right) = \hat \theta \left( t \right) - \theta \left( t \right) = \hat \theta \left( t \right) - {\Theta _{\kappa \left( t \right)}}$ is demonstrated only experimentally, but it is not proved analytically. In turn, for the proposed law \eqref{eq24}, the global exponential convergence is strictly analytically proved within Theorem.}

\textcolor{black}{There are other modifications \cite{c34, c36} of the concurrent law \eqref{eq31} with different algorithms for stack purging and update of ${\cal R}{\rm{,\;}}{\varphi _k}{\rm{,\;}}{y_k}$. However, being compared to \eqref{eq24}, their disadvantages are the same and can be summarized as follows:}
\begin{itemize}
\item[--] \textcolor{black}{the exponential convergence of the identification error $\tilde \theta \left( t \right) = \hat \theta \left( t \right) - \theta \left( t \right)$ of the piecewise constant function $\theta \left( t \right) = {\Theta _{\kappa \left( t \right)}}$ is demonstrated experimentally, but not analytically;}
\item[--]  \textcolor{black}{the update of ${\cal R}{\rm{,\;}}{\varphi _k}{\rm{,\;}}{y_k}$ happens occasionally, and therefore the rate of convergence can become substantially low;}
\item[--] \textcolor{black}{ the algorithms to form and store the data stack use nontrivial and often ambiguous algorithmic procedures for off-line processing of measurable signals.}
\end{itemize}

\subsubsection{DREM + Efficient Learning}\label{sec424}

\textcolor{black}{To identify the PC parameters without application of the offline procedures of monitoring and discrete update of the data stack, an alternative approach, which is based on the procedure of dynamic regressor extension and mixing, has been proposed in \cite{c14}.}

\textcolor{black}{The following extension of the regression equation \eqref{eq8} is introduced:
\begin{eqnarray}\label{eq33}
\begin{array}{c}
\dot z\left( t \right) =  - l\left( t \right)z\left( t \right) + \varphi \left( t \right)y\left( t \right){\rm{,\;}}z\left( {t_i^ + } \right) = 0,\\
\dot \omega \left( t \right) =  - l\left( t \right)\omega \left( t \right) + \varphi \left( t \right){\varphi ^{\rm{T}}}\left( t \right){\rm{,\;}}\omega \left( {t_i^ + } \right) = 0,
\end{array}
\end{eqnarray}
where $l\left( t \right)$ is calculated as follows:
\begin{eqnarray}\label{eq34}
l\left( t \right) = \left\{ \begin{array}{l}
{l_0}{{,\;if\;}}{\textstyle{{2{\lambda _{\min }}\left( {\omega \left( t \right)} \right) - \lambda _{\min }^{UB} - \lambda _{\min }^{LB}} \over {\lambda _{\min }^{UB} - \lambda _{\min }^{LB}}}} \ge 1\\
{\textstyle{{{l_0}} \over 2}}\left( {{\textstyle{{2{\lambda _{\min }}\left( {\omega \left( t \right)} \right) - \lambda _{\min }^{UB} - \lambda _{\min }^{LB}} \over {\lambda _{\min }^{UB} - \lambda _{\min }^{LB}}}} + 1} \right){{,\;otherwise}}
\end{array} \right.{\rm{,}}
\end{eqnarray}
where $0 < \lambda _{\min }^{LB} \le {\lambda _{\min }}\left( {\omega \left( t \right)} \right) \le \lambda _{\min }^{UB}$ are the minimum eigenvalue and its lower and upper bounds, ${l_0} > 0$ is a scaling factor.}

\textcolor{black}{On the basis of \eqref{eq33} it is proposed to use the following identification law, which is based on the DREM procedure \cite{c17}:
\begin{eqnarray}\label{eq35}
\dot{ \hat{ \theta}} \left( t \right) =  - {\gamma _0}det \left\{ {\omega \left( t \right)} \right\}\left( {det \left\{ {\omega \left( t \right)} \right\}\hat \theta \left( t \right) - adj\left\{ {\omega \left( t \right)} \right\}z\left( t \right)} \right){\rm{,\;}}{\gamma _0} > 0.
\end{eqnarray}}
\textcolor{black}{It is proved \cite{c14} that, owing to the implication ${\lambda _{\min }}\left( {\omega \left( t \right)} \right) \to \lambda _{\min }^{LB} \Rightarrow l\left( t \right) \to 0$, the filtering \eqref{eq33}, first of all, guarantees that $\omega \left( t \right)$ is globally bounded away from zero from some time instant $\bar t$, which can be determined using \linebreak ${\lambda _{\min }}\left( {\omega \left( t \right)} \right) = \lambda _{\min }^{LB} \Rightarrow \bar t = t$. On the other hand, according to \eqref{eq34}, if ${\lambda _{\min }}\left( {\omega \left( t \right)} \right) \to \lambda _{\min }^{UB}$, then $l\left( t \right) \to {l_0}$, and the filtering \eqref{eq33} has considerable sensitivity to changes of the unknown parameters values, i.e. the implication \linebreak ${\lambda _{\min }}\left( {\omega \left( t \right)} \right) \to \lambda _{\min }^{UB} \Rightarrow z\left( t \right) \to \omega \left( t \right)\theta \left( t \right)$ holds. An informal condition for high-quality tracking $\hat \theta \left( t \right)$ of the piecewise constant function $\theta \left( t \right)$ is a finite excitation of the regressor after each change of the regression parameters \eqref{eq2} and an increase of the eigenvalue ${\lambda _{\min }}\left( {\omega \left( t \right)} \right)$ to a predefined bound $ \lambda _{\min }^{UB}$ (a sufficient level of the regressor excitation) after a time instant $t_i^ + $.}

\textcolor{black}{The exponential convergence of the identification error $\tilde \theta \left( t \right) = \hat \theta \left( t \right) - \theta \left( t \right)$ for the law \eqref{eq35} is demonstrated in \cite{c14} only within a numerical experiment, which is its main disadvantage compared to the proposed law \eqref{eq24}. On the other hand, an important advantage of \eqref{eq35} over \eqref{eq24} is that it does not require to detect switching time instants $t_i^ + $. However, this is achieved at the expense of necessity to choose the values of $\lambda _{\min }^{LB}{\rm{,\;}}\lambda _{\min }^{UB}{\rm{,\;}}{l_0}$, from which the quality of the $\hat \theta \left( t \right)$ transients depends significantly.}
\section{Numerical experiments}\label{sec5}

To prove the efficiency of the developed procedure of piecewise constant unknown parameters identification, in this section the results of numerical experiments are presented. They were conducted in Matlab/Simulink using numerical integration with the help of the Euler method with the constant step size ${\tau _s} = {10^{ - 4}}$ seconds.

Section 5.1 presents the results of parameters identification of the regression equation \eqref{eq2} with and without the external disturbance. Section 5.2 is to show the results of identification of switched system parameters.

\subsection{Simple Example}\label{sec51}

\subsubsection{Noise Free Scenario}\label{sec51}

The regression \eqref{eq2} was chosen as:
\begin{eqnarray}\label{eq36}
\begin{array}{c}
\forall t \ge 0,{\rm{}\;}y\left(t\right) = \left[ {\begin{array}{*{20}{c}}
1&{{e^{ - t}}}
\end{array}} \right]{\Theta _{\kappa \left( t  \right)}}{\rm{,}}\;\;\kappa \left( t  \right) \in \left\{ {1,{\rm{ 2}}} \right\}{\rm{,}}\\
{\Theta _1} = {\left[ {\begin{array}{*{20}{c}}
{ - 2}&1
\end{array}} \right]^{\rm{T}}}{\rm{,}}\;{\Theta _2} = {\left[ {\begin{array}{*{20}{c}}
{ - 4}&2
\end{array}} \right]^{\rm{T}}}{\rm{,}}
\end{array}
\end{eqnarray}
where $\kappa \left( t \right)$ was to define the following switching sequence:
\begin{eqnarray}\label{eq37}
\begin{array}{c}
\Sigma  = \left\{ {\left( {1,{\rm{\;}}0} \right){\rm{,\;}}\left( {2,{\rm{\;}}0.5} \right){\rm{,}}\;\left( {1,{\rm{}}\;1} \right)} \right\}{\rm{,}}\\
{\Theta _{\kappa \left( t \right)}}{\rm{:}} = \left\{ \begin{array}{l}
{\Theta _1}{\rm{,}}\;\begin{array}{*{20}{c}}
{\forall t \in \left[ {0{\rm{; 0}}{\rm{.5}}} \right){\rm{,}}}\\
{\forall t \ge 1,}
\end{array}{\rm{}}\;\\
{\Theta _2}{\rm{,}}\;\forall t \in \left[ {0.{\rm{5; 1}}} \right){\rm{,}}
\end{array} \right.{\rm{,}}\left[ {{\rm{}}\;\begin{array}{*{20}{c}}
{{\theta _0} = {\Theta _1}}\\
{{\theta _1} = {\Theta _2}}\\
{{\theta _2} = {\Theta _1}}
\end{array}} \right]{\rm{.}}
\end{array}
\end{eqnarray}

The regression \eqref{eq36}, \eqref{eq37} met all requirements of Assumption 1. As we considered the disturbance free scenario, then the algorithm \eqref{eq16} was used to identify the elements of $\Im$. The parameters of filters \eqref{eq9}, \eqref{eq22a}-\eqref{eq22c}, estimation algorithm \eqref{eq16} and adaptive law \eqref{eq24} were chosen as:
\begin{eqnarray}\label{eq38}
\sigma  = 5,{\rm{}}\;{\Delta _{pr}} = 0.1,{\rm{}}\;k = 100, \; \rho  = {10^{ - 19}}{\rm{,}}\;{\gamma _0} = 10.
\end{eqnarray}

Figure 2 demonstrates the obtained transients of the regressor $\Omega \left( t \right)$ and residual $\epsilon \left( t \right)$.
\begin{figure}[h]
\centerline{\includegraphics[height=18pc]{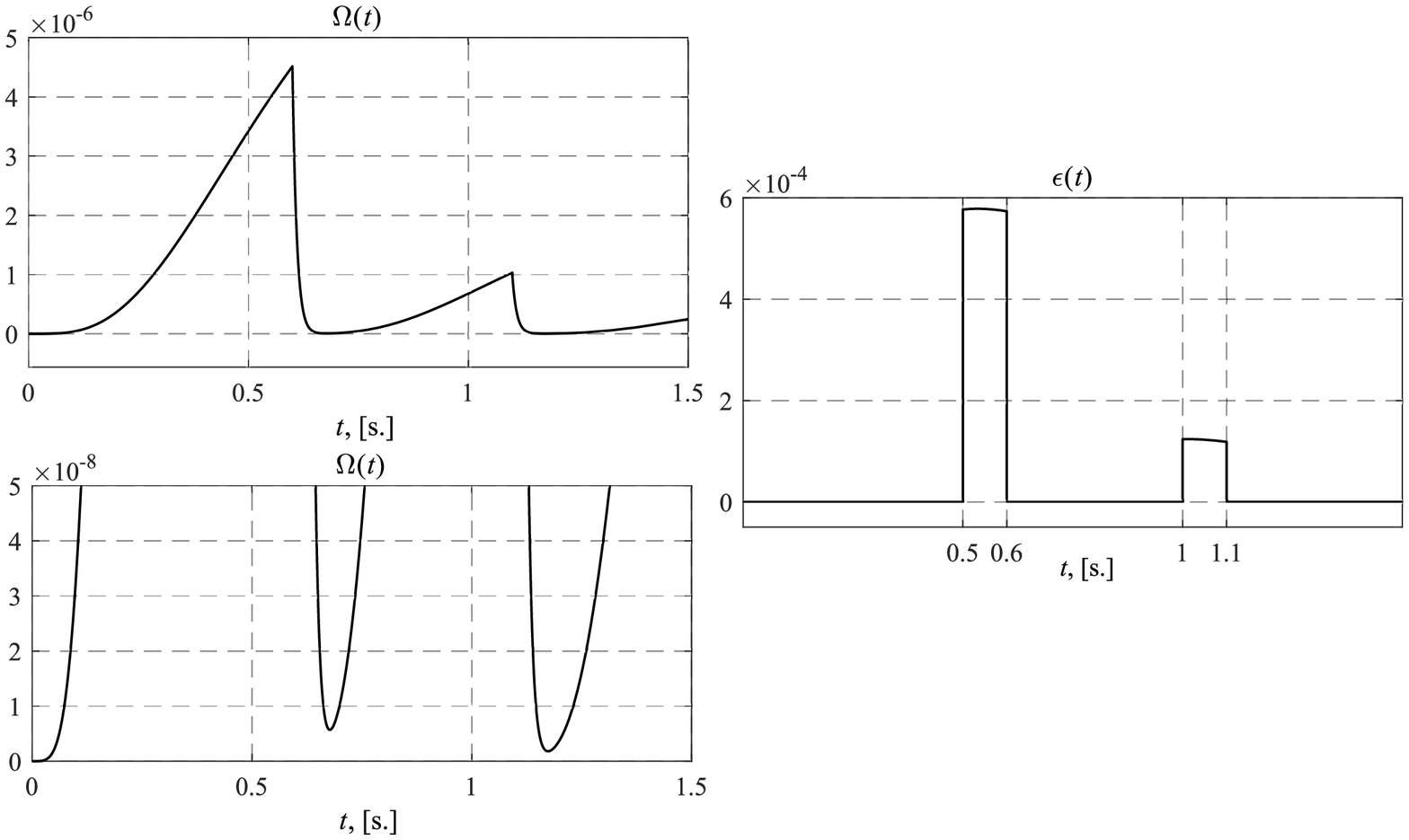}}
\caption{Transients of $\Omega \left( t \right)$ and $\epsilon \left( t \right)$\label{fig2}}
\end{figure}

The results, which are shown in Figure 2, corroborated the theoretical conclusions, which were made in Propositions 1 and 4. The regressor $\Omega \left( t \right)$ was bounded from below $\Omega \left( t \right) \ge {\Omega _{LB}} > 0$ for $t \ge t_0^ + {\rm{ + }}{T_0}$, and $\epsilon\left( t \right)$ was equal to zero $\forall t \in \left[ {\hat t_i^ + {\rm{;\;}}t_{i + 1}^ + } \right)$.

Figure 3 depicts the comparison of: 1) $t_i^+$ and its estimate $\hat t_i^ +$ obtained with the help of the algorithm \eqref{eq16}, and \linebreak 2) $\theta \left( t \right)$ and its estimate $\hat \theta \left( t \right)$, which was calculated using the adaptive law \eqref{eq24}.
\begin{figure}[h]
\centerline{\includegraphics[height=12pc]{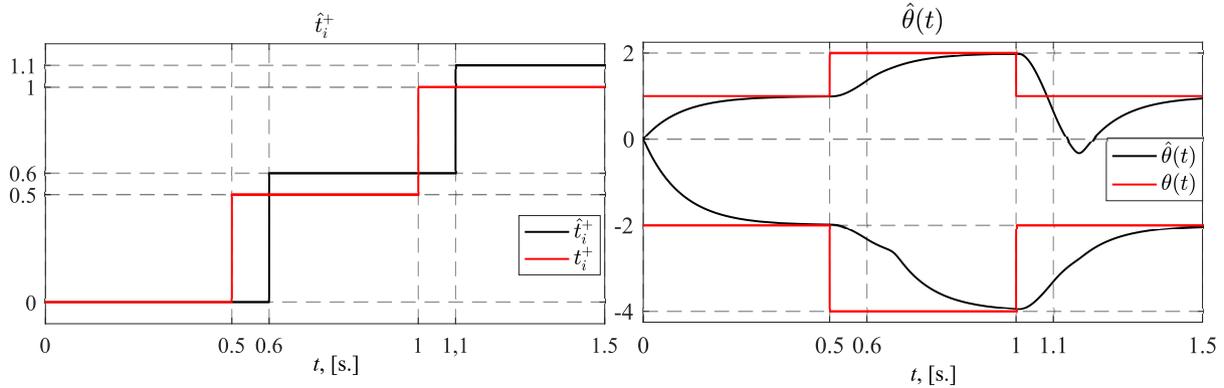}}
\caption{Transients of $\hat t_i^+$ and $\hat \theta \left( t \right)$ \label{fig3}}
\end{figure}

The transients of $\hat t_i^+$ and $\hat \theta \left( t \right)$ demonstrated all properties shown theoretically in Proposition 2 and Theorem. The dependence of the value of the estimation error $\tilde t_i^+$ from the value of ${\Delta _{pr}}$ was demonstrated. The error $\tilde \theta \left( t \right)$ converged to zero exponentially $\forall t \ge t_0^ + {\rm{ + }}{T_0}$.

Transients of $\hat t_i^+$ and $\hat \theta \left( t \right)$ for different values of ${\Delta _{pr}}$ and $\sigma$ respectively are depicted in Figure 4.
\begin{figure}[h]
\centerline{\includegraphics[height=12pc]{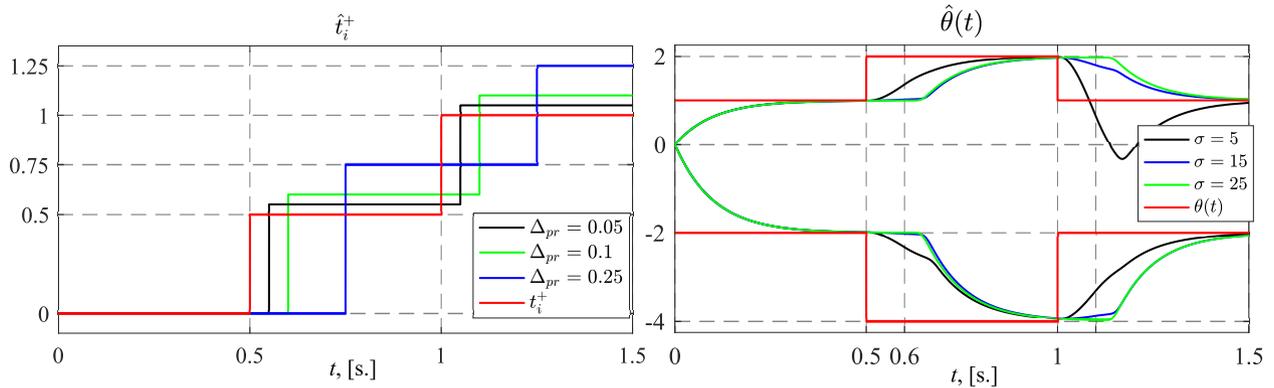}}
\caption{Transient curves of $\hat t_i^ + \left( {{\Delta _{pr}}} \right)$ and $\hat \theta \left( \sigma  \right)$ \label{fig4}}
\end{figure}

The obtained results shown in Fig. 4 confirmed that the error $\tilde t_i^+$ value could be adjusted by the choice of the parameter ${\Delta _{pr}}$. Also, they demonstrated that, as it was noted in Section 4, the quality of $\hat \theta \left( t \right)$ transients can be improved by choice of $\sigma$ value.

Then the proposed adaptive law was compared with the laws \eqref{eq30} and \eqref{eq35}, the parameters of which were set as follows:
\begin{eqnarray}\label{eq39}
\begin{array}{c}
{l_0} = 100,{\rm{\;}}\lambda _{\min }^{LB} = {10^{ - 6}}{\rm{,\;}}\lambda _{\min }^{UB} = {10^{ - 3}}{\rm{,\;}}{\gamma _0} = {10^{11}}{\rm{,\;}}{\Gamma _1} = 100{I_{2 \times 2}}{\rm{,\;}}{\Gamma _2} = 5000{I_{2 \times 2}}{\rm{,\;}}\\
\kappa \left( t \right) = 1 \Rightarrow {\varphi _1} = \varphi \left( {0.{\rm{05}}} \right){\rm{,\;}}{y_1} = y\left( {0.{\rm{05}}} \right){\rm{,\;}}{\varphi _2} = \varphi \left( {0.{\rm{1}}} \right){\rm{,\;}}{y_2} = y\left( {0.{\rm{1}}} \right){\rm{,\;}}\ell  = 2,\\
\kappa \left( t \right) = 2 \Rightarrow {\varphi _1} = \varphi \left( {0.{\rm{55}}} \right){\rm{,\;}}{y_1} = y\left( {0.{\rm{55}}} \right){\rm{,\;}}{\varphi _2} = \varphi \left( {0.{\rm{6}}} \right){\rm{,\;}}{y_2} = y\left( {0.{\rm{6}}} \right){\rm{,\;}}\ell  = 2.
\end{array}
\end{eqnarray}

\textcolor{black}{Figure 5 demonstrates the transients of estimates $\hat \theta \left( t \right)$ and $ {\hat \Theta _{\kappa \left( t \right)}}\left( t \right)$ obtained as a result of application of the laws \eqref{eq35} and \eqref{eq30}.}
\begin{figure}[h]
\centerline{\includegraphics[height=11pc]{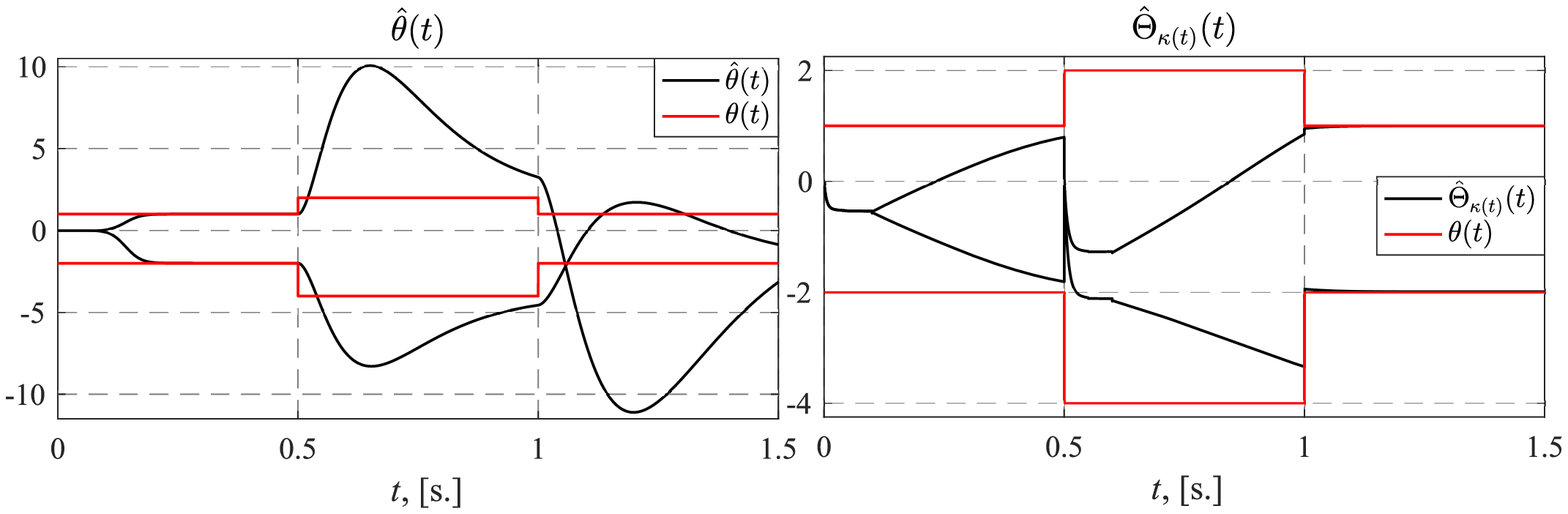}}
\caption{\textcolor{black}{Transient curves of $\hat \theta \left( t \right)$ for \eqref{eq35} and ${\hat \Theta _{\kappa \left( t \right)}}\left( t \right)$ for \eqref{eq30}} \label{fig4}}
\end{figure}
\textcolor{black}{Comparison of the transients in Fig. 5 with the ones shown in Fig. 4 and Fig. 3 made it possible to verify the advantages of the proposed solution. Compared to \eqref{eq35}, the law \eqref{eq24} ensured the exponential convergence of estimates $\hat \theta \left( t \right)$ to the function $\theta \left( t \right)$, and significant outliers did not occur in the course of the transients. Compared to \eqref{eq30}, the law \eqref{eq24} generated continuous estimates, had a higher rate of convergence, did not require storage and off-line processing of data on the regression equation \eqref{eq2}, and was sensitive to changes of the unknown parameters at the unknown time instants.}

Thus, the numerical experiments confirmed all the theoretically stated properties of the estimation algorithm \eqref{eq16} and the adaptive law \eqref{eq24}. The obtained transients of $\hat t_i^ +$ and $\hat \theta \left( t \right)$ indicated that all goals from \eqref{eq7} were achieved.

\subsubsection{Simple Noised Scenario}\label{sec412}

Then the estimation algorithm \eqref{eq19} and the adaptive law \eqref{eq24} were tested in the presence of an external disturbance caused by the measurement noise added to the regression equation \eqref{eq2}.
The regression \eqref{eq2} was chosen as:\textcolor{black}{
\begin{eqnarray}\label{eq40}
\begin{array}{c}
\forall t \ge 0,{\rm{\;}}y\left( t \right) = {\varphi ^{\rm{T}}}\left( t \right){\Theta _{\kappa \left( t \right)}}{\rm{ + }}w\left( t \right){\rm{,\;}}\kappa \left( t \right) \in \left\{ {1,{\rm{ 2}}} \right\}{\rm{,}}\\
{\varphi ^{\rm{T}}}\left( t \right) = \left[ {\begin{array}{*{20}{c}}
1&{{e^{ - t}}}
\end{array}} \right]{\rm{,\;}}w\left( t \right) = rand\left( {0.001} \right) - 0.5,
\end{array}
\end{eqnarray}
where the upper bound of $w\left( t \right)$ was considered to be known ${w_{\max }} = 0.65$. The signal $\kappa \left( t \right)$ was defined as the sequence \eqref{eq37}.}

The estimation algorithm \eqref{eq19} was used to identify the elements of the set $\Im$. The parameters of filters \eqref{eq9}, \eqref{eq22a}-\eqref{eq22c}, estimation algorithm \eqref{eq19} and adaptive law \eqref{eq24} were set as follows:
\begin{eqnarray}\label{eq41}
\begin{array}{c}
\sigma  = 25,{\rm{\;}}{\Delta _{pr}} = 0.01,{\rm{\;}}k = 100,{\rm{\;}}\rho  = 2.5 \cdot {10^{ - 11}}{\rm{,\;}}{\gamma _0} = 10,{\rm{\;}}\;
c\left( t \right) = E\left\{ {{w_{\max }}\varphi \left( t \right){\varphi ^{\rm{T}}}\left( t \right)adj\left\{ {\omega \left( t \right)} \right\}\int\limits_{\hat t_i^ + }^t {{e^{ - \int\limits_{\hat t_i^ +}^\tau  {\sigma ds} }}\varphi \left( \tau  \right)} {\rm{\;}}d\tau } \right\}.
\end{array}
\end{eqnarray}

Figure 6 is to show the comparison of a) ${\varphi ^{\rm{T}}}\left( t \right){\Theta _{\kappa \left( t \right)}}$ and $y\left( t \right)$; b) ${\rm E}\left\{ {\epsilon \left( t \right)} \right\}$ and $0.9\sqrt {{\mathop{\rm var}} \left\{ {\epsilon \left( t \right)} \right\}}  + c\left( t \right)$.
\vspace{-0.1cm}
\begin{figure}[h]
\centerline{\includegraphics[height=18pc]{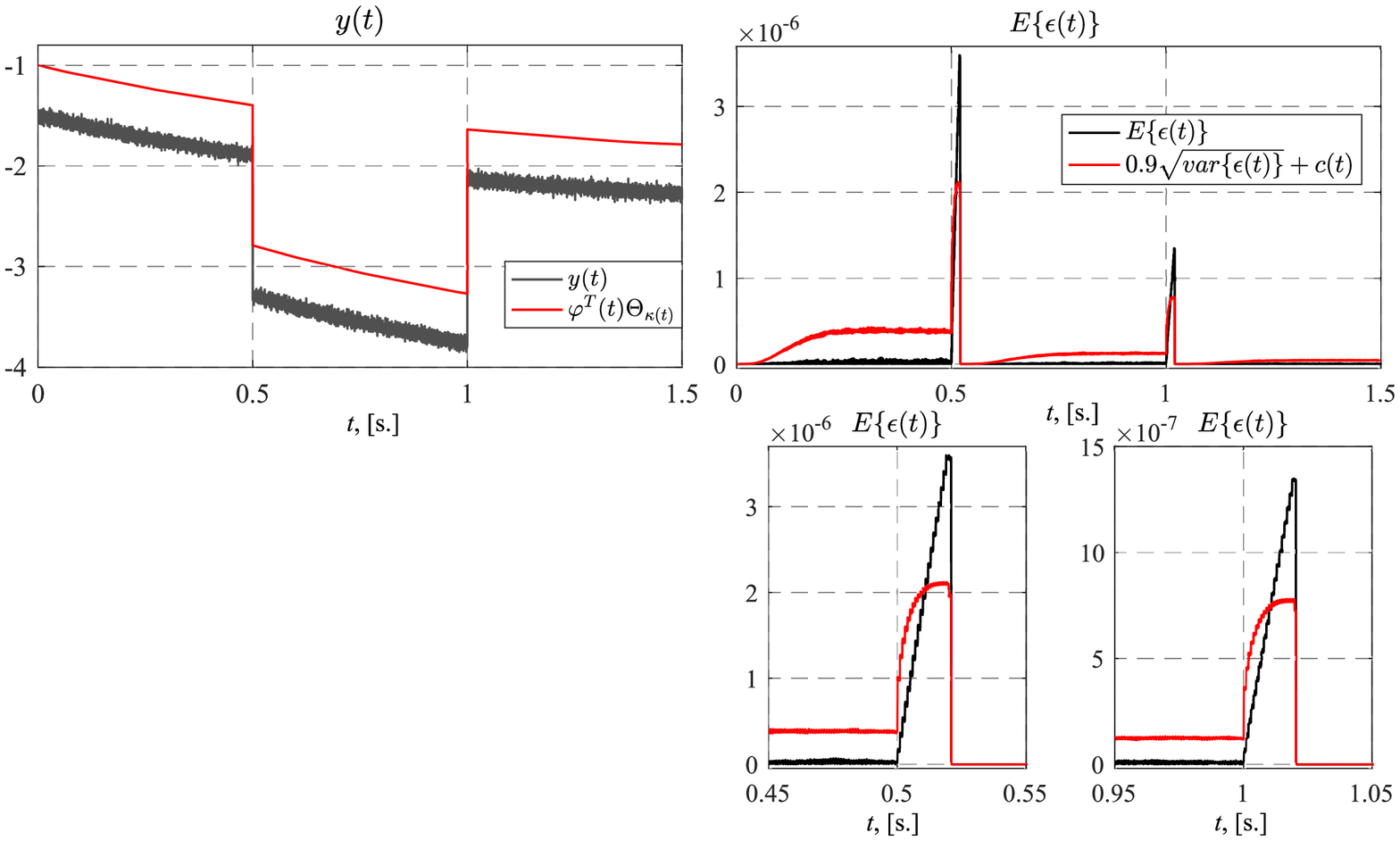}}
\caption{\textcolor{black}{Comparison of a) ${\varphi ^{\rm{T}}}\left( t \right){\Theta _{\kappa \left( t  \right)}}$ and $y\left( t \right)$ ; b) ${\rm E}\left\{ {\epsilon \left( t \right)} \right\}$ and $0.9\sqrt {{\mathop{\rm var}} \left\{ {\epsilon \left( t \right)} \right\}}  + c\left( t \right)$ \label{fig6}}}
\end{figure}

The results shown in Figure 6a demonstrated how the perturbation $w\left( t \right)$ affected the regression ${\varphi ^{\rm{T}}}\left( t \right){\Theta _{\kappa \left( \tau  \right)}}$. Figure 6b proves that the robust algorithm \eqref{eq19} accurately detected the time instants when the unknown parameters switched to their new values.

Figure 7 is to compare a) $t_i^+$ and $\hat t_i^+$ obtained using the algorithm \eqref{eq19}, and b) ${\theta(t)}$ and $\hat \theta \left( t \right)$ obtained with the help of the adaptive law \eqref{eq24}.
\begin{figure}[h]
\centerline{\includegraphics[height=12pc]{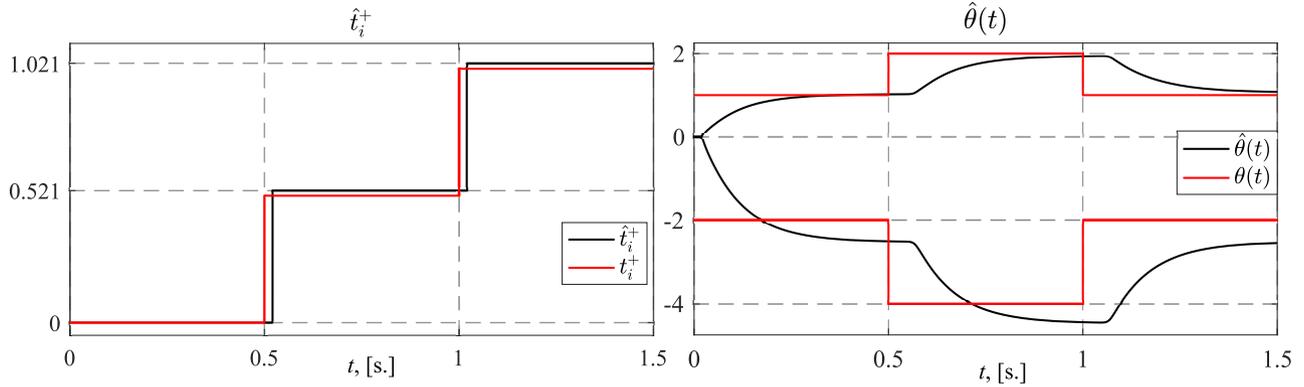}}
\caption{\textcolor{black}{Transients of $\hat t_i^+$ and $\hat \theta \left( t \right)$ \label{fig7}}}
\end{figure}

The results shown in Figure 7 confirmed that the robust algorithm \eqref{eq19} detected the elements of the set \eqref{eq6} in the presence of an external disturbance. The transients of $\hat \theta \left( t \right)$ demonstrated exponential convergence of the parameter error $\tilde \theta \left( t \right)$ to the compact set, which was described in Remark 5. 

Then the estimation algorithm \eqref{eq19} and the adaptive law \eqref{eq24} were tested in the presence of a disturbance, which was caused by both the measurement noise and a harmonic function added to the regression equation \eqref{eq2}. The regression and parameters were chosen as:
\begin{eqnarray}\label{eq42}
\begin{array}{c}
\forall t \ge 0,{\rm{\;}}y\left( t \right) = {\varphi ^{\rm{T}}}\left( t \right){\Theta _{\kappa \left( t \right)}}{\rm{ + }}w\left( t \right){\rm{,}}\;\kappa \left( t  \right) \in \left\{ {1,{\rm{ 2}}} \right\}{\rm{,}}\\
{\varphi ^{\rm{T}}}\left( t \right) = \left[ {\begin{array}{*{20}{c}}
1&{{e^{ - t}}}
\end{array}} \right]{\rm{,}}\;w\left( t \right) = rand\left( {0.001} \right){\rm{ + 0}}{\rm{.1}}sin\left( {25t} \right){\rm{,}}
\end{array}
\end{eqnarray}
where the upper bound of $w\left( t \right)$ was considered to be known ${w_{\max }} = 0.25$. The signal $\kappa \left( t \right)$ was used to define the switching sequence \eqref{eq28}.

The estimation algorithm \eqref{eq19} was used to detect the elements of the set $\Im$. The parameters of filters \eqref{eq9}, \eqref{eq22a}-\eqref{eq22c}, estimation algorithm \eqref{eq19} and adaptive law \eqref{eq24} were set in accordance with \eqref{eq41}.

Figure 8 demonstrates the comparison of (a) ${\varphi ^{\rm{T}}}\left( t \right){\Theta _{\kappa \left( t  \right)}}$ and $y\left( t \right)$; (b) ${\rm E}\left\{ {\epsilon \left( t \right)} \right\}$ and $0.9\sqrt {{\mathop{\rm var}} \left\{ {\epsilon \left( t \right)} \right\}}  + c\left( t \right)$.
\begin{figure}[h]
\centerline{\includegraphics[height=12pc]{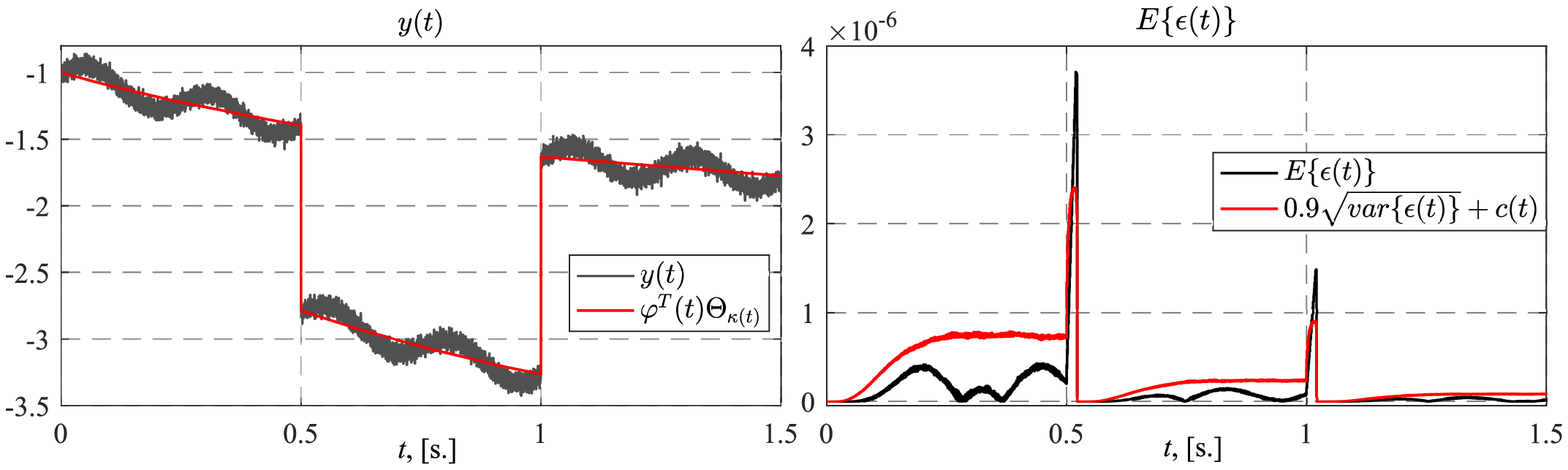}}
\caption{Comparison of (a) ${\varphi ^{\rm{T}}}\left( t \right){\Theta _{\kappa \left( t  \right)}}$ and $y\left( t \right)$; (b) ${\rm E}\left\{ {\epsilon \left( t \right)} \right\}$ and $0.9\sqrt {{\mathop{\rm var}} \left\{ {\epsilon \left( t \right)} \right\}}  + c\left( t \right)$ \label{fig8}}
\end{figure}

The results shown in Figure 8a demonstrates how the disturbance $w\left( t \right)$ affected the regression ${\varphi ^{\rm{T}}}\left( t \right){\Theta _{\kappa \left( t \right)}}$. Figure 8b proves correctness of the conclusions made in the propositions for the robust algorithm \eqref{eq19}.

Figure 9 shows the comparison of a) $t_i^ +$ and $\hat t_i^ + {\rm{,}}$ b) $\left\| {\tilde \theta \left( t \right)} \right\|$  and its asymptotic upper bound (UB) calculated as $UB = \left\| {W\left( t \right)} \right\|{\Omega ^{ - 1}}\left( t \right).$
\begin{figure}[h]
\centerline{\includegraphics[height=12pc]{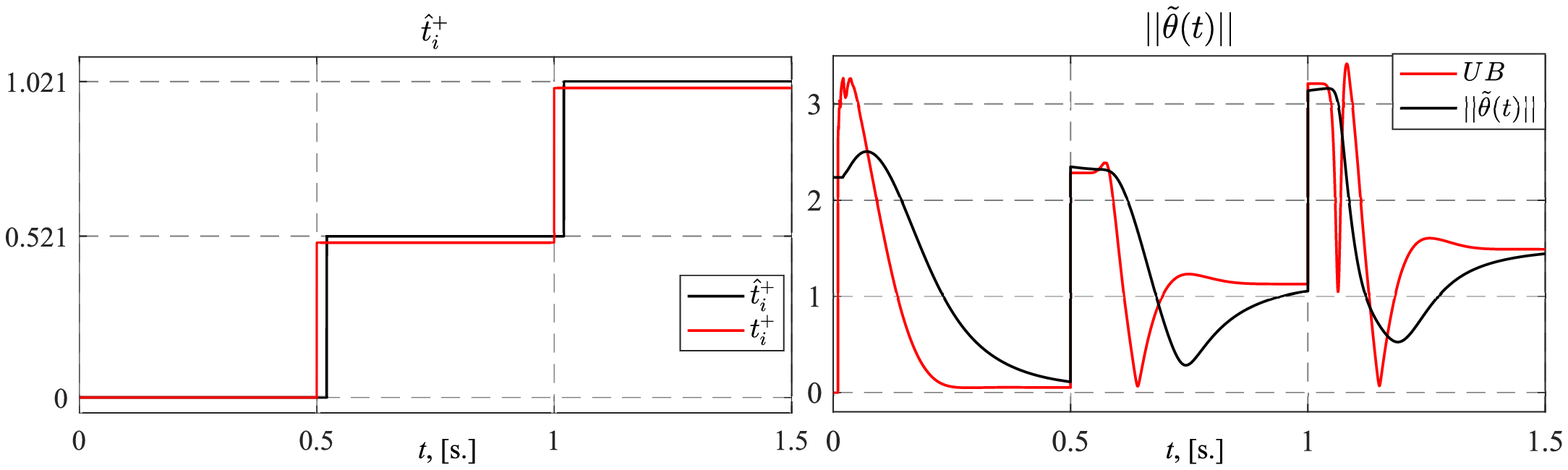}}
\caption{Transients of $\hat t_i^+$ and $\left\| {\tilde \theta \left( t \right)} \right\|$ \label{fig9}}
\end{figure}

The results shown in Figure 9 confirmed that the robust algorithm \eqref{eq19} detected the elements of the set \eqref{eq6} in case of a bounded disturbance. The transients of $\left\| {\tilde \theta \left( t \right)} \right\|$ converged exponentially to the compact set in full accordance with Remark 5. The value of UB over the intervals $\left[ {0.5{\rm{; 1}}} \right)$ and $\left[ {1{\rm{; 1}}{\rm{.5}}} \right)$ was substantially higher than the one over the time range $\left[ {0{\rm{;}}\;0.5} \right)$, because the level of the regressor $\varphi \left( t \right)$ excitation was vanishing with time and, as a consequence, the denominator of $UB = \left\| {W\left( t \right)} \right\|{\Omega ^{ - 1}}\left( t \right)$ significantly decreased.

The transients of the estimates $\hat \theta \left( t \right)$ obtained using different values of $\sigma$ are shown in Figure 10.
\begin{figure}[h]
\centerline{\includegraphics[height=14pc]{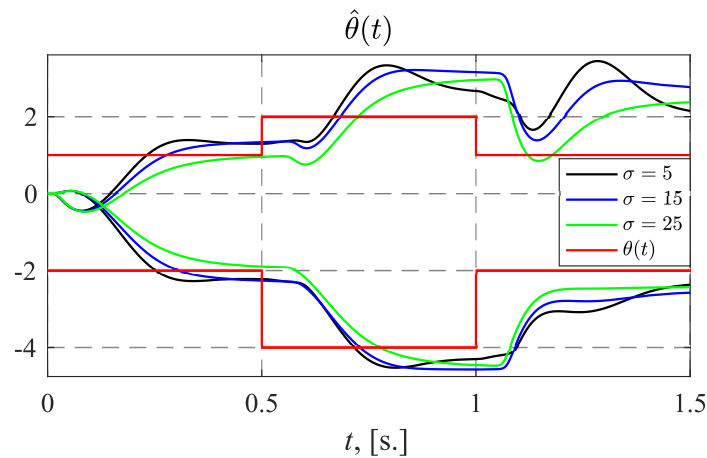}}
\caption{Transients of $\hat \theta \left( t \right)$ for different values of $\sigma$ \label{fig9}}
\end{figure}

It follows from Fig. 10 that, in the presence of an external disturbance $w\left( t \right)$, it is possible to improve the transient quality of the estimates $\hat \theta \left( t \right)$ by choice of the filter \eqref{eq9} parameter $\sigma$, which proved the recommendations given in Section 4.

Thus, the conducted experiments demonstrated the performance and efficiency of the developed estimation algorithm \eqref{eq19} and adaptive law \eqref{eq24} when the regression equation \eqref{eq2} is subjected to external bounded disturbances.

\subsection{Switched System Identification}\label{sec52}

In this subsection it is shown how the developed procedure can be applied to identify switched system parameters.

The following plant is considered:
\begin{eqnarray}\label{eq43}
\begin{array}{c}
\forall t \ge t_0^ + {\rm{,}}\;\dot x\left( t \right) = {{\overline \Theta^{\rm{T}}  _{\kappa \left( t \right)}}}\Phi\left( t \right) = {A_{\kappa \left( t \right)}}x\left( t \right) + {B_{\kappa \left( t \right)}}u\left( t \right){\rm{,}}\;x\left( {t_0^ + } \right) = {x_0}{\rm{,}}\\
\Phi \left( t \right) = {\left[ {\begin{array}{*{20}{c}}
{{x^{\rm{T}}}\left( t \right)}&{{u^{\rm{T}}}\left( t \right)}
\end{array}} \right]^{\rm{T}}}{\rm{,}}\;\overline \Theta  _{\kappa \left( t \right)}^{\rm{T}} = \left[ {\begin{array}{*{20}{c}}
{{A_{\kappa \left( t \right)}}}&{{B_{\kappa \left( t \right)}}}
\end{array}} \right]{\rm{,}}
\end{array}
\end{eqnarray}
where $x\left( t \right) \in {\mathbb{R}^n}$ is the state vector with the initial condition $x_0$, $u\left( t \right) \in {\mathbb{R}^m}$ is the control vector, ${A_{\kappa \left( t \right)}} \in {\mathbb{R}^{n \times n}}$ is the unknown state matrix, ${B_{\kappa \left( t \right)}} \in {\mathbb{R}^{n \times m}}$ is the unknown input matrix, $\kappa \left( t \right) \in \Xi  = \left\{ {1,2, \ldots ,N} \right\}$ is the unknown discrete function, which defines the switching time instants. The pair $\left( {{A_{\kappa \left( t \right)}}{\rm{}}, \;{B_{\kappa \left( t \right)}}} \right)$ is considered to be controllable, the vector $\Phi \left( t \right) \in {\mathbb{R}^{n + m}}$ is measurable  $\forall t > t_0^+$, and the parameter matrix ${\rm{}}\overline \Theta  _{\kappa \left( t \right)}^{\rm{T}} \in {\mathbb{R}^{n \times \left( {n + m} \right)}}$ is unknown $\forall t \ge t_0^+$.

It is assumed that the control signal $u\left( t \right)$ for \eqref{eq43} is formed by the following law:
\begin{eqnarray}\label{eq44}
u\left( t \right) = {K_x}x\left( t \right) + {K_r}r\left( t \right){\rm{,}}
\end{eqnarray}
where ${K_x} \in {\mathbb{R}^{m \times n}}$ is the matrix of feedback gains, ${K_r} \in {\mathbb{R}^{m \times m}}$ is the matrix of feedforward gains, $r\left( t \right) \in {\mathbb{R}^m}$ is the reference signal.

The derivative of the state vector $\dot x\left( t \right)$ is considered to be unknown. So, to represent the plant \eqref{eq43} as a linear regression \eqref{eq2} with measurable function $y\left(t\right)$, the filtration procedure on the basis of a stable filter is applied to \eqref{eq43}:
\begin{eqnarray}\label{eq45}
\begin{array}{c}
y\left( t \right) = x^{\rm{T}}\left( t \right) - l\overline x^{\rm{T}}\left( t \right) = {\varphi ^{\rm{T}}}\left( t \right){\Theta _{\kappa \left( t \right)}}{\rm{ + }}{w_{\kappa \left( t \right)}}{\rm{,}}\\
\varphi \left( t \right) = {\left[ {\begin{array}{*{20}{c}}
{{{\overline \Phi }^{\rm{T}}}\left( t \right)}&{{e^{ - l\left( {t - \hat t_i^ + } \right)}}}
\end{array}} \right]^{\rm{T}}}{\rm{,}}\;\Theta _{\kappa \left( t \right)}^{\rm{T}} = \left[ {\begin{array}{*{20}{c}}
{{A_{\kappa \left( t \right)}}}&{{B_{\kappa \left( t \right)}}}&{x\left( {\hat t_i^ + } \right)}
\end{array}} \right]{\rm{,}}\\
\dot {\overline \Phi} \left( t \right) =  - l\overline \Phi \left( t \right) + \Phi \left( t \right){\rm{,}}\;\overline \Phi \left( {\hat t_i^ + } \right) = {0_{m + n}}{\rm{,}}
\end{array}
\end{eqnarray}
where $l > 0$ is the filter constant, $\overline \Phi \left( t \right) \in {\mathbb{R}^{m + n}}$ is the filtered regressor, $\overline x\left( t \right) \in {\mathbb{R}^n}$ is an element of the vector $\overline \Phi \left( t \right)$, \linebreak $\Theta _{\kappa \left( t \right)}^{\rm{T}} \in {\mathbb{R}^{n \times \left( {m + n + 1} \right)}}$ is the extended vector of the unknown parameters, $\varphi \left( t \right) \in {\mathbb{R}^{m + n + 1}}$ is the extended regressor vector, ${w_{\kappa \left( t \right)}} \in {\mathbb{R}^{1 \times n}}$ is the disturbance, which is caused by the plant \eqref{eq43} parameters switch and inertial characteristics of the filter from \eqref{eq45}. According to the recommendations given in Remark 4, as far as the parameterization \eqref{eq45} is considered, the filter is reset at time points $\hat t_i^+$. More details on how to obtain \eqref{eq45} from \eqref{eq43} can be found in \cite{c5, c8,c9,c10}.

We assumed that $\kappa \left( t \right) \in \Xi  = \left\{ {1,{\rm{ 2}}} \right\}$ and the matrices ${A_{\kappa \left( t \right)}}{\rm{,}}\;{B_{\kappa \left( t \right)}}$ and $x\left( {t_0^ + } \right)$ were defined as follows:
\begin{eqnarray}\label{eq46}
{A_1} = \left[ {\begin{array}{*{20}{c}}
0&1\\
{ - 6}&{ - 8}
\end{array}} \right]{\rm{,}}\;{B_1} = \left[ {\begin{array}{*{20}{c}}
0\\
2
\end{array}} \right]{\rm{,}}\;{A_2} = \left[ {\begin{array}{*{20}{c}}
0&1\\
{ - 2}&{ - 4}
\end{array}} \right]{\rm{,}}\;{B_2} = \left[ {\begin{array}{*{20}{c}}
0\\
4
\end{array}} \right]{\rm{,}}\;{x_0} = \left[ {\begin{array}{*{20}{c}}
{ - 1}\\
0
\end{array}} \right]{\rm{,}}
\end{eqnarray}
and $\kappa \left( t \right)$ was to represent the following switching sequence:
\begin{eqnarray}\label{eq47}
\Sigma  = \left\{ {\left( {1,{\rm{\;0}}} \right){\rm{,}}\;\left( {{\rm{2}}{\rm{,\;5}}} \right){\rm{,}}\;\left( {1,{\rm{\;10}}} \right)} \right\}.
\end{eqnarray}

The disturbance ${w_{\kappa \left( t \right)}}$ was caused by the plant \eqref{eq43} parameters switch and $\forall t \in \left[ {\hat t_i^ + {\rm{;}}\;t_{i + 1}^ + } \right){\rm{}}\;{w_{\kappa \left( t \right)}} = 0$, whereas \linebreak $\forall t \in \left[ {t_i^ + {\rm{;}}\;\hat t_i^ + } \right){\rm{}}\;{w_{\kappa \left( t \right)}} \ne 0$. It is easy to prove the correctness of the above equations by analogy with the proof of Propositions 1 and 3. Therefore, in spite of the disturbance ${w_{\kappa \left( t \right)}}$ in \eqref{eq45}, the estimation algorithm \eqref{eq16} was applied to identify the sequence $\Im$ in the course of the experiment.

The parameters of the control law \eqref{eq44}, filters \eqref{eq9}, \eqref{eq22a}-\eqref{eq22c}, \eqref{eq45}, estimation algorithm \eqref{eq16}, and adaptive law \eqref{eq24} were chosen as follows:
\begin{eqnarray}\label{eq48}
{K_x}{\rm{ = }}\left[ {\begin{array}{*{20}{c}}
{ - 5}&{ - 4}
\end{array}} \right]{\rm{,}}\;{K_r} = 8,{\rm{}}\;r\left( t \right) = 1,{\rm{}}\;\sigma  = 5,{\rm{}}\;k = 100, \; {\Delta _{pr}} = 0.1,{\rm{}}\;\rho  = {10^{ - 17}}{\rm{,}}\;{\gamma _0} = 10.
\end{eqnarray}

Figure 11 presents the comparison of $t_i^+$ and estimate $\hat t_i^+$ obtained with the help of algorithm \eqref{eq16}, as well as comparison of parameters ${\theta (t)}$ and estimates $\hat \theta \left( t \right)$ calculated with the help of the adaptive law \eqref{eq24}.
\begin{figure}[h]
\centerline{\includegraphics[height=20pc]{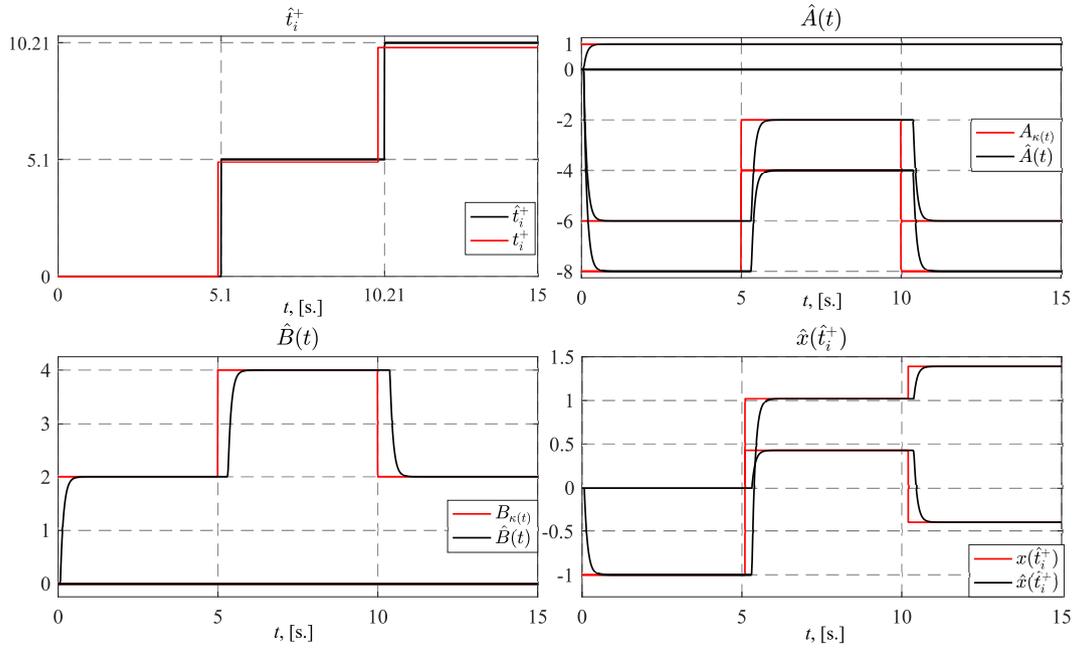}}
\caption{Transients of $\hat t_i^+$ and $\hat \theta \left( t \right)$ \label{fig11}}
\end{figure}

The shown transients demonstrates that the goal \eqref{eq7} was achieved, and the estimation algorithm \eqref{eq16} and the adaptive law \eqref{eq24} could be used to identify the parameters of the switched system \eqref{eq43}.

\textcolor{black}{The identification law \eqref{eq24} was compared with the classical ones used to identify the parameters of linear switched systems \eqref{eq43}. For this purpose, the adjustable models were introduced into consideration:
\begin{eqnarray}\label{eq49}
\forall t \ge t_0^ + {\rm{,\;}}\left\{ \begin{array}{l}
{{\dot {\hat{ x}}}_j}\left( t \right) = {A_m}{{\hat x}_j}\left( t \right) + \left( {{{\hat A}_j}\left( t \right) - {A_m}} \right)x\left( t \right) + {{\hat B}_i}\left( t \right)u\left( t \right){\rm{,\;}}if{\rm{\;}}\kappa \left( t \right) = j\\
{{\hat x}_j}\left( t \right) = x\left( t \right){\rm{,\;}}if{\rm{\;}}\kappa \left( t \right) \ne j
\end{array} \right.
\end{eqnarray}
Taking into consideration \eqref{eq49}, the Lyapunov-based laws were introduced:
\begin{eqnarray}\label{eq50}
{\dot {\hat{ A}}_j}\left( t \right) = \left\{ \begin{array}{l}
 - \Gamma _A^jP{{\tilde x}_j}\left( t \right){x^{\rm{T}}}\left( t \right){\rm{,\;if\;}}\kappa \left( t \right) = j,\\
0,{\rm{\;if\;}}\kappa \left( t \right) \ne j,
\end{array} \right.{\rm{\;}}{\dot {\hat{ B}}_j}\left( t \right) = \left\{ \begin{array}{l}
 - \Gamma _B^jP{{\tilde x}_j}\left( t \right){u^{\rm{T}}}\left( t \right){\rm{,\;if\;}}\kappa \left( t \right) = j,\\
0,{\rm{\;if\;}}\kappa \left( t \right) \ne j,
\end{array} \right.
\end{eqnarray}
as well as their composite modifications:
\begin{eqnarray}\label{eq51}
\begin{array}{l}
{{\dot {\hat{ A}}}_j}\left( t \right) = \left\{ \begin{array}{l}
 - \Gamma _A^jP{{\tilde x}_j}\left( t \right){x^{\rm{T}}}\left( t \right) - \bar \Gamma _A^j\sum\limits_{k = 1}^\ell  {\left( {{{\hat A}_j}\left( t \right){x_k} + {{\hat B}_j}\left( t \right){u_k} - {{\dot x}_k}} \right)x_k^{\rm{T}}} {\rm{,\;if\;}}\kappa \left( t \right) = j\\
 - \bar \Gamma _A^j\sum\limits_{k = 1}^\ell  {\left( {{{\hat A}_j}\left( t \right){x_k} + {{\hat B}_j}\left( t \right){u_k} - {{\dot x}_k}} \right)x_k^{\rm{T}}} {\rm{,\;if\;}}\kappa \left( t \right) \ne j
\end{array} \right.{\rm{,\;}}\\
{{\dot {\hat {B}}}_j}\left( t \right) = \left\{ \begin{array}{l}
 - \Gamma _B^jP{{\tilde x}_j}\left( t \right){u^{\rm{T}}}\left( t \right) - \bar \Gamma _B^j\sum\limits_{k = 1}^\ell  {\left( {{{\hat A}_j}\left( t \right){x_k} + {{\hat B}_j}\left( t \right){u_k} - {{\dot x}_k}} \right)u_k^{\rm{T}}} {\rm{,\;if\;}}\kappa \left( t \right) = j\\
 - \bar \Gamma _B^j\sum\limits_{k = 1}^\ell  {\left( {{{\hat A}_j}\left( t \right){x_k} + {{\hat B}_j}\left( t \right){u_k} - {{\dot x}_k}} \right)u_k^{\rm{T}}} {\rm{,\;if\;}}\kappa \left( t \right) \ne j
\end{array} \right.,
\end{array}
\end{eqnarray}
where ${A_m}$ was a Hurwitz matrix, $P$ was a solution of the Lyapunov equation $A_m^{\rm{T}}P + P{A_m} =  - Q$, the derivative ${\dot x_k} = \dot x\left( {{t_k}} \right)$ value was measured using Optimal Point Smoother \cite{c6}, which provided reasonable accuracy.}

\textcolor{black}{The parameters of the laws \eqref{eq50} and \eqref{eq51} were chosen as follows:
\begin{eqnarray}\label{eq52}
\begin{array}{c}
\Gamma _A^1 = \Gamma _B^1 = 0.025{I_2}{\rm{,\;}}\bar \Gamma _A^1 = 160{I_2}{\rm{,\;}}\bar \Gamma _B^1 = 3.5{I_2}{\rm{,\;}}\Gamma _A^2 = \Gamma _B^2 = 0.25{I_2}{\rm{,\;}}\bar \Gamma _A^2 = 100{I_2}{\rm{,\;}}\bar \Gamma _B^2 = 20.95{I_2}{\rm{,}}\\
\kappa \left( t \right) = 1 \Rightarrow {t_k} \in \left\{ {0.05,{\rm{ 0}}{\rm{.1}}{\rm{, 0}}{\rm{.15}}} \right\}{\rm{,\;}}\kappa \left( t \right) = 2 \Rightarrow {t_k} \in \left\{ {0.505,{\rm{ 0}}{\rm{.51}}{\rm{, 0}}{\rm{.515}}} \right\}
\end{array}
\end{eqnarray}}

\textcolor{black}{Figure 12 depicts the transients of $\hat \theta \left( t \right)$ obtained as a result of application of the laws \eqref{eq50} and \eqref{eq51} respectively.}

\begin{figure}[h]
\centerline{\includegraphics[height=20pc]{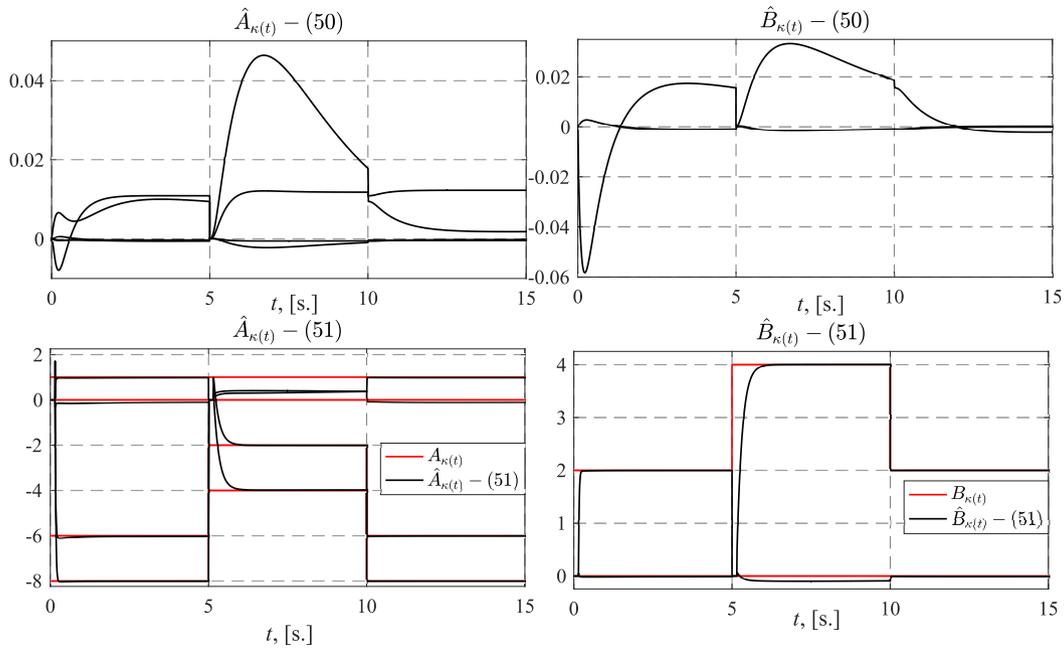}}
\caption{\textcolor{black}{Transients of $\hat \theta \left( t \right)$ obtained as a result of application of the laws \eqref{eq50} and \eqref{eq51}} \label{fig11}}
\end{figure}

\textcolor{black}{Comparing the transients in Fig. 11 and Fig. 12, it was possible to observe the main differences between \eqref{eq24} and \eqref{eq50} and \eqref{eq51}, as well as to verify the advantages of the proposed solution \eqref{eq24}. Namely, the law \eqref{eq24} formed continuous estimates of piecewise constant parameters, did not require {\it{a priori}} knowledge or dynamic estimation of the function $\kappa \left( t \right)$.}

Thus, the experiments have demonstrated that the estimation algorithm \eqref{eq16} or \eqref{eq19} and adaptive law \eqref{eq24} can be applied to solve various identification and adaptive control problems, which can be reduced to the identification problem of unknown piecewise constant parameters of the regression equation \eqref{eq2}.

\section{Conclusion}\label{sec5}

\textcolor{black}{Given that the switching sequence was unknown, a procedure with only one adaptive law to track piecewise constant unknown parameters of a linear regression equation was proposed. It provided: 1) adjustable accuracy of estimates of the switching time instants without necessity of data stacks offline processing, 2) global exponential convergence of the unknown PC parameters estimates to their true values if the regressor was finitely exciting somewhere inside the time interval between two consecutive changes of the parameters.}

\textcolor{black}{Robustness of the proposed identification procedure to influence of the external disturbances was analytically proved. Recommendations on the choice of the arbitrary parameters of the procedure were given to improve the quality of transients of the unknown parameters estimates for both cases: with and without external disturbances.}

\textcolor{black}{All theoretically proved properties of the developed identification procedure were experimentally confirmed. Particularly, the procedure was applied to estimate the parameters of the linear switched system.}

\textcolor{black}{The scope of future research is to use the proposed procedure to develop adaptive control method for switched systems, which ensures a global exponential stability of a closed-loop system under condition that a regressor is finitely exciting after each parameters switch, and requires neither any knowledge of a plant input matrix, nor the switching time instants.}


\section*{Acknowledgments}
This research was financially supported by Grants Council of the President of the Russian Federation (project MD-1787.2022.4). 

\subsection*{Conflict of interest}

The authors declare no potential conflict of interests.

\appendix

\section{Proof of Proposition 1\label{app1}}

As $\kappa \left( t \right) =j_i$  holds $\forall t \in \left[ {\hat t_i^ + {\rm{;}}\;t_{i + 1}^ + } \right)$ when $\hat t_i^ +  \ge t_i^ +$, then the unknown parameters of the regressions \eqref{eq8} and \eqref{eq11} coincide to each other $\forall t \in \left[ {\hat t_i^ + {\rm{;}}\;t_{i + 1}^ + } \right)$. In its turn, according to \eqref{eq9}, it is obtained:

\begin{eqnarray}\label{eq_a01}
\forall t \in \left[ {\hat t_i^ + {\rm{;\;}}t_{i + 1}^ + } \right){\rm{,\;}}z\left( t \right) = \int\limits_{\hat t_i^ + }^t {{e^{ - \int\limits_{\hat t_i^ + }^\tau  {\sigma ds} }}\varphi \left( \tau  \right)y\left( \tau  \right)} {\rm{\;}}d\tau  = \int\limits_{\hat t_i^ + }^t {{e^{ - \int\limits_{\hat t_i^ + }^\tau  {\sigma ds} }}\varphi \left( \tau  \right){\varphi ^{\rm{T}}}\left( \tau  \right)} {\rm{\;}}d\tau {\theta _i} = \omega \left( t \right){\theta _i}.
\end{eqnarray}

The function \eqref{eq_a01} is substituted into \eqref{eq14} to obtain:
\begin{eqnarray}\label{eq_a02}
\forall t \in \left[ {\hat t_i^ + {\rm{;\;}}t_{i + 1}^ + } \right){\rm{\;}} \epsilon \left( t \right){\rm{:}} = \varphi \left( t \right){\varphi ^{\rm{T}}}\left( t \right)adj\left\{ {\omega \left( t \right)} \right\}\omega \left( t \right){\theta _i} - \Delta \left( t \right)\varphi \left( t \right){\varphi ^{\rm{T}}}\left( t \right){\theta _i} = 0_{n \times p},
\end{eqnarray}
as it was to be proved for the first time interval.

So, the next time range is considered: $\left[ {t_i^ + {\rm{;}}\;\hat t_i^ + } \right)$. The equation for the function $z\left( t \right)$ from \eqref{eq9} is rewritten in the following form:
\begin{eqnarray}\label{eq_a03}
\begin{array}{l}
\forall t \in \left[ {t_i^ + {\rm{;\;}}\hat t_i^ + } \right){\rm{,\;}}z\left( t \right) = \int\limits_{\hat t_{i - 1}^ + }^t {{e^{ - \int\limits_{\hat t_{i - 1}^ + }^\tau  {\sigma ds} }}\varphi \left( \tau  \right)y\left( \tau  \right)} {\rm{\;}}d\tau  = \int\limits_{\hat t_{i - 1}^ + }^{t_i^ + } {{e^{ - \int\limits_{\hat t_{i - 1}^ + }^\tau  {\sigma ds} }}\varphi \left( \tau  \right){\varphi ^{\rm{T}}}\left( \tau  \right)} {\rm{\;}}d\tau {\theta _{i - 1}} 
 + \int\limits_{t_i^ + }^t {{e^{ - \int\limits_{\hat t_{i - 1}^ + }^\tau  {\sigma ds} }}\varphi \left( \tau  \right){\varphi ^{\rm{T}}}\left( \tau  \right)} {\rm{\;}}d\tau {\theta _i}.
\end{array} 
\end{eqnarray}

Then \eqref{eq_a03} is substituted into \eqref{eq14} to obtain:
\begin{eqnarray}\label{eq_a04}
\begin{array}{c}
\forall t \in \left[ {t_i^ + {\rm{;\;}}\hat t_i^ + } \right){\rm{,\;}} \epsilon \left( t \right){\rm{:}} = \varphi \left( t \right){\varphi ^{\rm{T}}}\left( t \right)adj\left\{ {\omega \left( t \right)} \right\}\int\limits_{\hat t_{i - 1}^ + }^{t_i^ + } {{e^{ - \int\limits_{\hat t_{i - 1}^ + }^\tau  {\sigma ds} }}\varphi \left( \tau  \right){\varphi ^{\rm{T}}}\left( \tau  \right)} {\rm{\;}}d\tau {\theta _{i - 1}} + \\
 + \varphi \left( t \right){\varphi ^{\rm{T}}}\left( t \right)adj\left\{ {\omega \left( t \right)} \right\}\int\limits_{t_i^ + }^t {{e^{ - \int\limits_{\hat t_{i - 1}^ + }^\tau  {\sigma ds} }}\varphi \left( \tau  \right){\varphi ^{\rm{T}}}\left( \tau  \right)} {\rm{\;}}d\tau {\theta _i} - \Delta \left( t \right)\varphi \left( t \right)y\left( t \right) \pm \\
 \pm \varphi \left( t \right){\varphi ^{\rm{T}}}\left( t \right)adj\left\{ {\omega \left( t \right)} \right\}\int\limits_{\hat t_{i - 1}^ + }^{t_i^ + } {{e^{ - \int\limits_{\hat t_{i - 1}^ + }^\tau  {\sigma ds} }}\varphi \left( \tau  \right){\varphi ^{\rm{T}}}\left( \tau  \right)} {\rm{\;}}d\tau {\theta _i} 
 = \varphi \left( t \right){\varphi ^{\rm{T}}}\left( t \right)adj\left\{ {\omega \left( t \right)} \right\}\int\limits_{\hat t_{i - 1}^ + }^{t_i^ + } {{e^{ - \int\limits_{\hat t_{i - 1}^ + }^\tau  {\sigma ds} }}\varphi \left( \tau  \right){\varphi ^{\rm{T}}}\left( \tau  \right)} {\rm{\;}}d\tau \left( {{\theta _{i - 1}} - {\theta _i}} \right){\rm{,}}
\end{array}
\end{eqnarray}
as it was to be proved for the second time interval.

The combination of \eqref{eq_a02} and \eqref{eq_a04} is the equation \eqref{eq15}, which completes the proof of Proposition 1.

\section{Proof of Proposition 2\label{app2}}
Following Proposition 2 statement and proof of Proposition 1, it is written:
\begin{eqnarray}\label{eq_b05}
\forall t \ge t_i^ + {\rm{,}} \left\| \epsilon {\left( t \right)} \right\| = \left\| {\varphi \left( t \right){\varphi ^{\rm{T}}}\left( t \right)adj\left\{ {\omega \left( t \right)} \right\}\int\limits_{\hat t_{i - 1}^ + }^{t_i^ + } {{e^{ - \int\limits_{\hat t_{i - 1}^ + }^\tau  {\sigma ds} }}\varphi \left( \tau  \right){\varphi ^{\rm{T}}}\left( \tau  \right)} {\rm{\;}}d\tau \left( {{\theta _{i - 1}} - {\theta _i}} \right)} \right\|.
\end{eqnarray}

According to Problem Statement section, Assumption 1 and Definition 1 (eq. \eqref{eq1}), the lower bounds for the second and third multipliers of \eqref{eq_b05} are obtained:
\begin{eqnarray}\label{eq_b06}
\begin{array}{l}
\int\limits_{\hat t_{i - 1}^ + }^{t_i^ + } {{e^{ - \int\limits_{\hat t_{i - 1}^ + }^\tau  {\sigma ds} }}\varphi \left( \tau  \right){\varphi ^{\rm{T}}}\left( \tau  \right)} {\rm{\;}}d\tau 
 = \int\limits_{\hat t_{i - 1}^ + }^{t_{i - 1}^ +  + {T_{i - 1}}} {{e^{ - \int\limits_{\hat t_{i - 1}^ + }^\tau  {\sigma ds} }}\varphi \left( \tau  \right){\varphi ^{\rm{T}}}\left( \tau  \right)d\tau }  + \int\limits_{t_{i - 1}^ +  + {T_{i - 1}}}^{t_i^ + } {{e^{ - \int\limits_{\hat t_{i - 1}^ + }^\tau  {\sigma ds} }}\varphi \left( \tau  \right){\varphi ^{\rm{T}}}\left( \tau  \right)d\tau }  \ge \\
 \ge \int\limits_{\hat t_{i - 1}^ + }^{t_{i - 1}^ +  + {T_{i - 1}}} {{e^{ - \int\limits_{\hat t_{i - 1}^ + }^\tau  {\sigma ds} }}\varphi \left( \tau  \right){\varphi ^{\rm{T}}}\left( \tau  \right)d\tau }  \ge {e^{ - \sigma \left( { - \tilde t_{i - 1}^ +  + {T_{i - 1}}} \right)}}\int\limits_{\hat t_{i - 1}^ + }^{t_{i - 1}^ +  + {T_{i - 1}}} {\varphi \left( \tau  \right){\varphi ^{\rm{T}}}\left( \tau  \right)d\tau } 
 \ge {{\bar \alpha }_{i - 1}}{e^{ - \sigma \left( { - \tilde t_{i - 1}^ +  + {T_{i - 1}}} \right)}}I_{n \times n} > 0
\end{array}
\end{eqnarray}

\begin{eqnarray}\label{eq_b07}
\begin{array}{l}
adj\left\{ {\omega \left( t \right)} \right\} = adj\left\{ {\int\limits_{\hat t_{i - 1}^ + }^t {{e^{ - \int\limits_{\hat t_{i - 1}^ + }^\tau  {\sigma ds} }}\varphi \left( \tau  \right){\varphi ^{\rm{T}}}\left( \tau  \right)d\tau } } \right\} = \\
 = adj\left\{ \begin{array}{l}
\int\limits_{\hat t_{i - 1}^ + }^{t_{i - 1}^ +  + {T_{i - 1}}} {{e^{ - \int\limits_{\hat t_{i - 1}^ + }^\tau  {\sigma ds} }}\varphi \left( \tau  \right){\varphi ^{\rm{T}}}\left( \tau  \right)d\tau }  + \int\limits_{t_{i - 1}^ +  + {T_{i - 1}}}^{t_i^ + } {{e^{ - \int\limits_{\hat t_{i - 1}^ + }^\tau  {\sigma ds} }}\varphi \left( \tau  \right){\varphi ^{\rm{T}}}\left( \tau  \right)d\tau }  + \\
 + \int\limits_{t_i^ + }^{t_i^ +  + {T_i}} {{e^{ - \int\limits_{\hat t_{i - 1}^ + }^\tau  {\sigma ds} }}\varphi \left( \tau  \right){\varphi ^{\rm{T}}}\left( \tau  \right)d\tau }  + \int\limits_{t_i^ +  + {T_i}}^t {{e^{ - \int\limits_{\hat t_{i - 1}^ + }^\tau  {\sigma ds} }}\varphi \left( \tau  \right){\varphi ^{\rm{T}}}\left( \tau  \right)d\tau } 
\end{array} \right\} \ge \\
 \ge adj\left\{ {\int\limits_{\hat t_{i - 1}^ + }^{t_{i - 1}^ +  + {T_{i - 1}}} {{e^{ - \int\limits_{\hat t_{i - 1}^ + }^\tau  {\sigma ds} }}\varphi \left( \tau  \right){\varphi ^{\rm{T}}}\left( \tau  \right)d\tau } } \right\} \ge adj\left\{ {{{\bar \alpha }_{i - 1}}{e^{ - \sigma \left( { - \tilde t_{i - 1}^ +  + {T_{i - 1}}} \right)}}I_{n \times n}} \right\} > 0
\end{array}
\end{eqnarray}

As $\varphi \left( t \right) \in {\rm{FE}}$ over the time range $\left[ {t_i^ + {\rm{;}}\;t_i^ +  + {T_i}} \right]$, then $\exists \left[ {{t_a}{\rm{;}}\;{t_b}} \right] \subset \left[ {t_i^ + {\rm{;}}\;t_i^ +  + {T_i}} \right]$ such that:
\begin{eqnarray}\label{eq_b08}
\forall t \in \left[ {{t_a}{\rm{;}}\;{t_b}} \right]{\rm{}}\; {\varphi \left( t \right){\varphi ^{\rm{T}}}\left( t \right)} > 0_{n \times n}
\end{eqnarray}
in the sense of positive definiteness.

The equations \eqref{eq_b06},\eqref{eq_b07},\eqref{eq_b08} are substituted into \eqref{eq_b05} to obtain $\forall t \in \left[ {{t_a}{\rm{;}}\;{t_b}} \right] \subset \left[ {t_i^ + {\rm{;}}\;t_i^ +  + {T_i}} \right]{\rm{}}\;\left\| {\varepsilon \left( t \right)} \right\| > 0$. Then, according to the algorithm \eqref{eq16}, for a general case it is obtained that ${t_{up}} = {t_c} \in \left[ {{t_a}{\rm{;}}\;{t_b}} \right]$ and $\hat t_i^ +  = {t_c} + {\Delta _{pr}}$. As $\tilde t_i^ +  = \hat t_i^ +  - t_i^ + $, then:
\begin{eqnarray}\label{eq_b09}
\tilde t_i^ +  = {t_c} + {\Delta _{pr}} - t_i^ +.
\end{eqnarray}

Taking into consideration ${t_c} \in \left[ {{t_a}{\rm{;}}\;{t_b}} \right] \subset \left[ {t_i^ + {\rm{;}}\;t_i^ +  + {T_i}} \right]$, the following is obtained from \eqref{eq_b09}:
\begin{eqnarray}\label{eq_b10}
{\Delta _{pr}} \le \tilde t_i^ +  \le {T_i} + {\Delta _{pr}},
\end{eqnarray}

It is concluded from \eqref{eq_b10} that the fact that $\hat t_i^ +  \ge t_i^+$ and appropriate choice of ${\Delta _{pr}}$ value ensure achievement of $\tilde t_i^ +  \le {T_i}$.

\section{Proof of Proposition 3\label{app3}}
Considering the relation (expression) $\epsilon \left( t \right) = \varphi \left( t \right){\varphi ^{\rm{T}}}\left( t \right)\varepsilon \left( t \right)$, the interval-based definition \eqref{eq21} of the disturbance $\varepsilon \left( t \right)$ follows directly from Proposition 1.

The following notations are introduced to obtain the upper bound of the disturbance $\varepsilon \left( t \right)$:
\begin{eqnarray}\label{eq_c13}
\begin{array}{c}
\forall t \in \left[ {t_i^ + {\rm{;\;}}\hat t_i^ + } \right),{\rm{\;}}\omega \left( t \right) = \int\limits_{\hat t_{i - 1}^ + }^t {{e^{ - \int\limits_{\hat t_{i - 1}^ + }^\tau  {\sigma ds} }}\varphi \left( \tau  \right){\varphi ^{\rm{T}}}\left( \tau  \right)} {\rm{\;}}d\tau  = {\omega _1}\left( t \right) + {\omega _2}\left( t \right){\rm{,}}\\
{\omega _1}\left( t \right) = \int\limits_{\hat t_{i - 1}^ + }^{t_i^ + } {{e^{ - \int\limits_{\hat t_{i - 1}^ + }^\tau  {\sigma ds} }}\varphi \left( \tau  \right){\varphi ^{\rm{T}}}\left( \tau  \right)} {\rm{\;}}d\tau ,{\rm{\;}}{\omega _2}\left( t \right) = \int\limits_{t_i^ + }^t {{e^{ - \int\limits_{\hat t_{i - 1}^ + }^\tau  {\sigma ds} }}\varphi \left( \tau  \right){\varphi ^{\rm{T}}}\left( \tau  \right){\rm{\;}}} d\tau .
\end{array}
\end{eqnarray}

The following estimates are correct for the exponentially decaying multipliers of the integrands from \eqref{eq_c13}:
\begin{eqnarray}\label{eq_c14}
\begin{array}{l}
\forall \tau  \in \left[ {\hat t_{i - 1}^ + {\rm{;\;}}t_i^ + } \right]{\rm{:\;}}{e^{ - \sigma \left( {t_i^ +  - \hat t_{i - 1}^ + } \right)}} \le {e^{ - \int\limits_{\hat t_{i - 1}^ + }^\tau  {\sigma ds} }} \le 1,\\
\forall \tau  \in \left[ {t_i^ + {\rm{;\;}}\hat t_i^ + } \right]{\rm{:\;}}{e^{ - \sigma \left( {\hat t_i^ +  - \hat t_{i - 1}^ + } \right)}} \le {e^{ - \int\limits_{\hat t_{i - 1}^ + }^\tau  {\sigma ds} }} \le {e^{ - \sigma \left( {t_i^ +  - \hat t_{i - 1}^ + } \right)}}.
\end{array}
\end{eqnarray}

Considering $\varphi \left( t \right) \in {L_\infty }$, \eqref{eq_c14} and the mean value theorem, the upper bounds of ${\omega _1}\left( t \right)$ and ${\omega _2}\left( t \right)$ are obtained:
\begin{eqnarray}\label{eq_c15}
\begin{array}{l}
{\omega _1}\left( t \right) = \int\limits_{\hat t_{i - 1}^ + }^{t_i^ + } {{e^{ - \int\limits_{\hat t_{i - 1}^ + }^\tau  {\sigma ds} }}\varphi \left( \tau  \right){\varphi ^{\rm{T}}}\left( \tau  \right)} {\rm{\;}}d\tau  \le \int\limits_{\hat t_{i - 1}^ + }^{t_i^ + } {\varphi \left( \tau  \right){\varphi ^{\rm{T}}}\left( \tau  \right)} {\rm{\;}}d\tau  \le {{\bar \delta }_1}\left( {t_i^ +  - \hat t_{i - 1}^ + } \right)I_{n \times n}{\rm{,}}\\
{{\bar \delta }_1} = {\rm{ess}}\mathop {{\rm{sup}}}\limits_{\hat t_{i - 1}^ +  \le t \le t_i^ + } {\lambda _{\max }}\left( {\varphi \left( t \right){\varphi ^{\rm{T}}}\left( t \right)} \right){\rm{,}}\\
{\omega _2}\left( t \right) = \int\limits_{t_i^ + }^t {{e^{ - \int\limits_{\hat t_{i - 1}^ + }^\tau  {\sigma ds} }}\varphi \left( \tau  \right){\varphi ^{\rm{T}}}\left( \tau  \right){\rm{\;}}} d\tau  \le {e^{ - \sigma \left( {t_i^ +  - \hat t_{i - 1}^ + } \right)}}\int\limits_{t_i^ + }^{\hat t_i^ + } {\varphi \left( \tau  \right){\varphi ^{\rm{T}}}\left( \tau  \right){\rm{\;}}} d\tau  \le {e^{ - \sigma \left( {t_i^ +  - \hat t_{i - 1}^ + } \right)}}{{\bar \delta }_2}\left( {\hat t_i^ +  - t_i^ + } \right)I_{n \times n}{\rm{,}}\\
{{\bar \delta }_2} = {\rm{ess}}\mathop {{\rm{sup}}}\limits_{t_i^ +  \le t \le \hat t_i^ + } {\lambda _{\max }}\left( {\varphi \left( t \right){\varphi ^{\rm{T}}}\left( t \right)} \right){\rm{,}}
\end{array}
\end{eqnarray}

The equality $adj\left\{ {A + B} \right\} = adj\left\{ A \right\} + adj\left\{ B \right\}{\rm{\;}}\forall A,B$ is applied to the definition of $\varepsilon \left( t \right)$. As a result:
\begin{eqnarray}\label{eq_c16}
\varepsilon \left( t \right) = adj\left\{ {\omega \left( t \right)} \right\}{\omega _1}\left( {{\theta _{i - 1}} - {\theta _i}} \right) = \left( {adj\left\{ {{\omega _1}\left( t \right)} \right\}{\omega _1}\left( t \right) + adj\left\{ {{\omega _2}\left( t \right)} \right\}{\omega _1}\left( t \right)} \right)\underbrace {\left( {{\theta _{i - 1}} - {\theta _i}} \right)}_{{\Delta _\theta }}.
\end{eqnarray}

The upper bounds from \eqref{eq_c15} are substituted into \eqref{eq_c16}. Then $adj\left\{ {c{I_{n \times n}}} \right\}\!=\! c \cdot adj\left\{ {{I_{n \times n}}} \right\}{\rm{,}}\;det \left\{ {c{I_{n \times n}}} \right\} = {c^n} \cdot det \left\{ {{I_{n \times n}}} \right\}{\rm{}}\;\forall c$ are applied to the obtained result. Finally, we obtain:
\begin{eqnarray}\label{eq_c17}
\left\| {\varepsilon \left( t \right)} \right\| \le \underbrace {\left( {\bar \delta _1^n{{\left( {t_i^ +  - \hat t_{i - 1}^ + } \right)}^n} + {{\bar \delta }_1}{{\bar \delta }_2}\left( {t_i^ +  - \hat t_{i - 1}^ + } \right)\left( {\hat t_i^ +  - t_i^ + } \right){e^{ - \sigma \left( {t_i^ +  - \hat t_{i - 1}^ + } \right)}}} \right){\Delta _\theta }}_{{\varepsilon _{max}}}{\rm{.}}
\end{eqnarray}

It follows from \eqref{eq_c17} that the disturbance $\varepsilon \left( t \right)$ is bounded.

The final aim is to analyze the regressor $\Delta \left( t \right)$ properties. The first step is, considering $\Delta \left( t \right) = det \left\{ {\omega \left( t \right)} \right\}$, to obtain the regressor $\omega \left( t \right)$ lower bound.

To achieve this, the bounds of the exponentially decaying multiplier of the integrand from the definition of $\omega \left( t \right)$ is written:
\begin{eqnarray}\label{eq_c18}
\forall \tau  \in \left[ {\hat t_i^ + {\rm{;\;}}t_i^ +  + {T_i}} \right]{\rm{,\;}}{e^{ - \sigma \left( { - \tilde t_i^ +  + {T_i}} \right)}} \le {e^{ - \int\limits_{\hat t_i^ + }^\tau  {\sigma ds} }} \le 1.
\end{eqnarray}

Considering \eqref{eq_c18}, the mean value theorem and the requirements of Assumption 1, the lower and upper bounds of the regressor $\omega \left( t \right)$ are obtained:

\begin{eqnarray}\label{eq_c19}
\begin{array}{l}
\forall t \ge t_i^ +  + {T_i}{\rm{,\;}}\omega \left( t \right) = \int\limits_{\hat t_i^ + }^t {{e^{ - \int\limits_{\hat t_i^ + }^\tau  {\sigma ds} }}\varphi \left( \tau  \right){\varphi ^{\rm{T}}}\left( \tau  \right)} {\rm{\;}}d\tau  \le {\textstyle{{{{\bar \delta }_3}} \over \sigma }}I_{n \times n}\\
{{\bar \delta }_3} = {\rm{ess}}\mathop {{\rm{sup}}}\limits_{\hat t_i^ +  \le t \le t_{i + 1}^ + } {\lambda _{\max }}\left( {\varphi \left( t \right){\varphi ^{\rm{T}}}\left( t \right)} \right)\\
\forall t \ge t_i^ +  + {T_i}{\rm{,\;}}\omega \left( t \right) = \int\limits_{\hat t_i^ + }^t {{e^{ - \int\limits_{\hat t_i^ + }^\tau  {\sigma ds} }}\varphi \left( \tau  \right){\varphi ^{\rm{T}}}\left( \tau  \right)} {\rm{\;}}d\tau  = \\
 = \int\limits_{\hat t_i^ + }^{t_i^ +  + {T_i}} {{e^{ - \int\limits_{\hat t_i^ + }^\tau  {\sigma ds} }}\varphi \left( \tau  \right){\varphi ^{\rm{T}}}\left( \tau  \right)} {\rm{\;}}d\tau  + \int\limits_{t_i^ +  + {T_i}}^t {{e^{ - \int\limits_{\hat t_i^ + }^\tau  {\sigma ds} }}\varphi \left( \tau  \right){\varphi ^{\rm{T}}}\left( \tau  \right)} {\rm{\;}}d\tau  \ge \\
 \ge \int\limits_{\hat t_i^ + }^{t_i^ +  + {T_i}} {{e^{ - \int\limits_{\hat t_i^ + }^\tau  {\sigma ds} }}\varphi \left( \tau  \right){\varphi ^{\rm{T}}}\left( \tau  \right)} {\rm{\;}}d\tau  \ge {e^{ - \sigma \left( { - \tilde t_i^ +  + {T_i}} \right)}}\int\limits_{\hat t_i^ + }^{t_i^ +  + {T_i}} {\varphi \left( \tau  \right){\varphi ^{\rm{T}}}\left( \tau  \right)} {\rm{\;}}d\tau  \ge {{\bar \alpha }_i}{e^{ - \sigma \left( { - \tilde t_i^ +  + {T_i}} \right)}}I_{n \times n} > 0
\end{array}
\end{eqnarray}

Now this result is used to obtain bounds of the regressor $\Delta \left( t \right)$. As $det \left\{ {c{I_{n \times n}}} \right\} = {c^n} \cdot det \left\{ {{I_{n \times n}}} \right\}{\rm{}}\;\forall c$, then the lower bound of the regressor $\Delta \left( t \right)$ is:
\begin{eqnarray}\label{eq_c20}
\forall t \ge t_i^ + {\rm{ + }}{T_i}{\rm{,}}\underbrace {\mathop {\max }\limits_{\forall i} {{\left( {{\textstyle{{{{\bar \delta }_3}} \over \sigma }}} \right)}^n}}_{{\Delta _{UB}}} \ge {\left( {{\textstyle{{{{\bar \delta }_3}} \over \sigma }}} \right)^n} \ge \Delta \left( t \right) \ge {\left( {{{\bar \alpha }_i}{e^{ - \sigma \left( { - \tilde t_i^ +  + {T_i}} \right)}}} \right)^n} \ge \underbrace {\mathop {\min }\limits_{{{\bar \alpha }_i}{\rm{,}} - \tilde t_i^ +  + {T_i}} {{\left( {{{\bar \alpha }_i}{e^{ - \sigma \left( { - \tilde t_i^ +  + {T_i}} \right)}}} \right)}^n}}_{{\Delta _{LB}}} > 0,
\end{eqnarray}
which completes the proof of Proposition 3.

\section{Proof of Proposition 4 \label{app4}}
To prove the proposition, the solutions of the equation \eqref{eq22b} over the intervals $\left[ {\hat t_i^ + {\rm{;\;}}t_i^ +  + {T_i}} \right]$ and $\left[ {t_i^ +  + {T_i}{\rm{;\;}}\hat t_{i + 1}^ + } \right]$ are written:
\begin{eqnarray}\label{eq_d21}
\begin{array}{l}
\forall t \in \left[ {\hat t_i^ + {\rm{;\;}}t_i^ +  + {T_i}} \right]{\rm{\;}}\Omega \left( t \right) = {e^{ - k\left( {t - \hat t_i^ + } \right)}}\Omega \left( {\hat t_i^ + } \right) + \int\limits_{\hat t_i^ + }^t {{e^{ - k\left( {t - \tau } \right)}}\Delta \left( \tau  \right){\mathop{\rm d}\nolimits} \tau } \\
\forall t \in \left[ {t_i^ +  + {T_i}{\rm{;\;}}\hat t_{i + 1}^ + } \right]{\rm{\;}}\Omega \left( t \right) = {e^{ - k\left( {t - t_i^ +  - {T_i}} \right)}}\Omega \left( {t_i^ +  + {T_i}} \right) + \int\limits_{t_i^ +  + {T_i}}^t {{e^{ - k\left( {t - \tau } \right)}}} \Delta \left( \tau  \right){\mathop{\rm d}\nolimits} \tau .
\end{array}
\end{eqnarray}

Taking into consideration the proof of Proposition 3, the following equations hold for $\Delta \left( t \right)$ over the above-stated time ranges:
\begin{eqnarray}\label{eq_d22}
\begin{array}{c}
\forall t \in \left[ {\hat t_i^ + {\rm{;\;}}t_i^ +  + {T_i}} \right]{\rm{\;}}\Delta \left( t \right) \equiv 0,\\
\forall t \in \left[ {t_i^ +  + {T_i}{\rm{;\;}}\hat t_{i + 1}^ + } \right]{\rm{\;}}{\Delta _{UB}} \ge \Delta \left( t \right) \ge {\Delta _{LB}} > 0.
\end{array}
\end{eqnarray}

The equation \eqref{eq_d22} is substituted into \eqref{eq_d21} to obtain the following estimates of $\Omega \left( t \right)$:
\begin{eqnarray}\label{eq_d23}
\begin{array}{c}
\forall t \in \left[ {\hat t_0^ + {\rm{;\;}}t_0^ +  + {T_0}} \right]{\rm{\;}}\Omega \left( t \right) \equiv 0,\\
\forall i \ge 1{\rm{\;}}\forall t \in \left[ {\hat t_i^ + {\rm{;\;}}t_i^ +  + {T_i}} \right]{\rm{\;}}\Omega \left( {\hat t_i^ + } \right) \ge \Omega \left( t \right) \ge {e^{ - k{T_i}}}\Omega \left( {\hat t_i^ + } \right) > 0,\\
\forall t \in \left[ {t_i^ +  + {T_i}{\rm{;\;}}\hat t_{i + 1}^ + } \right]{\rm{\;}}\Omega \left( {t_i^ +  + {T_i}} \right) + \left( {\hat t_{i + 1}^ +  - t_i^ +  - {T_i}} \right){\Delta _{UB}} \ge \Omega \left( t \right) \ge \\
 \ge {e^{ - k\left( {\hat t_{i + 1}^ +  - t_i^ +  - {T_i}} \right)}}\Omega \left( {t_i^ +  + {T_i}} \right) + \left( {\hat t_{i + 1}^ +  - t_i^ +  - {T_i}} \right){\Delta _{LB}} > 0.
\end{array}
\end{eqnarray}

From \eqref{eq_d23} we have:
\begin{eqnarray}\label{eq_d24}
\begin{array}{c}
\forall t \ge t_0^ + {\rm{ + }}{T_0}{\rm{\;}}{\Omega _{{\rm{UB}}}} \ge \Omega \left( t \right) \ge {\Omega _{{\rm{LB}}}} > 0,\\
{\Omega _{{\rm{LB}}}} = \mathop {\min }\limits_{\forall i \ge 1} \left\{ {{e^{ - k\left( {\hat t_{i + 1}^ +  - t_i^ +  - {T_i}} \right)}}\Omega \left( {t_i^ +  + {T_i}} \right) + \left( {\hat t_{i + 1}^ +  - t_i^ +  - {T_i}} \right){\Delta _{LB}}{\rm{,\;}}{e^{ - k{T_i}}}\Omega \left( {\hat t_i^ + } \right)} \right\}{\rm{,}}\\
{\Omega _{{\rm{UB}}}} = \mathop {\max }\limits_{\forall i \ge 1} \left\{ {\Omega \left( {t_i^ +  + {T_i}} \right) + \left( {\hat t_{i + 1}^ +  - t_i^ +  - {T_i}} \right){\Delta _{UB}}{\rm{,\;}}\Omega \left( {\hat t_i^ + } \right)} \right\}{\rm{,}}
\end{array}
\end{eqnarray}
which completes the proof of Proposition 4.

\section{Proof of Theorem\label{app5}}
As the inequality $\Omega \left( t \right) > \rho $ holds, the differential equation, which describes the dynamics of the parameter error $\forall t \ge t_0^ + {\rm{ + }}{T_0}$, takes the form:
\begin{eqnarray}\label{eq_e25}
	\dot {\tilde \theta} \left( t \right) = \dot {\hat \theta} \left( t \right) - \theta \left( t \right) =  - \gamma {\Omega ^2}\left( t \right)\tilde \theta \left( t \right) + \gamma \Omega \left( t \right)d\left( t \right) - \dot \theta \left( t \right) =  - {\gamma _0}\tilde \theta \left( t \right) + {\gamma _0}{\textstyle{{d\left( t \right)} \over {\Omega \left( t \right)}}} - \dot \theta \left( t \right).
\end{eqnarray}

The solution of \eqref{eq_e25} is written as:
\begin{eqnarray}\label{eq_e26}
\begin{array}{l}
\tilde \theta \left( t \right) = {e^{ - {\gamma _0}\left( {t - t_0^ +  - {T_0}} \right)}}\tilde \theta \left( {t_0^ +  + {T_0}} \right) + {\gamma _0}\int\limits_{t_0^ +  + {T_0}}^t {{e^{ - {\gamma _0}\left( {t - \tau } \right)}}{\textstyle{{d\left( \tau  \right)} \over {\Omega \left( \tau  \right)}}}d\tau }  - \int\limits_{t_0^ +  + {T_0}}^t {{e^{ - {\gamma _0}\left( {t - \tau } \right)}}\dot \theta \left( \tau  \right)d\tau }  = \\
 = {e^{ - {\gamma _0}\left( {t - t_0^ +  - {T_0}} \right)}}\tilde \theta \left( {t_0^ +  + {T_0}} \right) + {\gamma _0}\int\limits_{t_0^ +  + {T_0}}^t {{e^{ - {\gamma _0}\left( {t - \tau } \right)}}{\textstyle{{d\left( \tau  \right)} \over {\Omega \left( \tau  \right)}}}d\tau }  - \int\limits_{t_0^ +  + {T_0}}^t {{e^{ - {\gamma _0}\left( {t - \tau } \right)}}\sum\limits_{q = 1}^i {\Delta _q^\theta \delta \left( {\tau  - t_q^ + } \right)} d\tau } .
\end{array}
\end{eqnarray}

Using the sifting property of the Dirac function:
\begin{eqnarray}\label{eq_e27}
	\int\limits_{t_0^ + }^t {f\left( \tau  \right)\delta \left( {\tau  - t_q^ + } \right)d\tau }  = f\left( {t_q^ + } \right)h\left( {t - t_q^ + } \right){\rm{,\;}}\forall f\left( t \right){\rm{,}}
\end{eqnarray}

it is obtained from \eqref{eq_e27}:
\begin{eqnarray}\label{eq_e28}
\begin{array}{c}
\tilde \theta \left( t \right) = {e^{ - {\gamma _0}\left( {t - t_0^ +  - {T_0}} \right)}}\tilde \theta \left( {t_0^ +  + {T_0}} \right) + {\gamma _0}\int\limits_{t_0^ +  + {T_0}}^t {{e^{ - {\gamma _0}\left( {t - \tau } \right)}}{\textstyle{{d\left( \tau  \right)} \over {\Omega \left( \tau  \right)}}}d\tau }  - \sum\limits_{q = 1}^i {{e^{ - {\gamma _0}\left( {t - t_q^ + } \right)}}\Delta _q^\theta \delta \left( {\tau  - t_q^ + } \right)}  = \\
 = \underbrace {\left( {\tilde \theta \left( {t_0^ +  + {T_0}} \right) - \sum\limits_{q = 1}^i {{e^{{\gamma _0}\left( {t_q^ +  - t_0^ +  - {T_0}} \right)}}\Delta _q^\theta h\left( {t - t_q^ + } \right)} } \right)}_{\beta \left( t \right)}{e^{ - {\gamma _0}\left( {t - t_0^ +  - {T_0}} \right)}} + {\gamma _0}\int\limits_{t_0^ +  + {T_0}}^t {{e^{ - {\gamma _0}\left( {t - \tau } \right)}}{\textstyle{{d\left( \tau  \right)} \over {\Omega \left( \tau  \right)}}}d\tau } .
\end{array}
\end{eqnarray}

The boundedness of the function $\beta \left( t \right)$ needs to be shown. If the statement 5.1 of Assumption 1 holds, then the number of switches of the system parameters is finite: $i \le {i_{{\rm{max}}}} < \infty $, and the following upper bound is written for $\beta \left( t \right)$:
\begin{eqnarray}\label{eq_e29}
	\left\| {\beta \left( t \right)} \right\| \le \left\| {\tilde \theta \left( {t_0^ +  + {T_0}} \right)} \right\| + \sum\limits_{q = 1}^{{i_{\max }}} {{e^{{\gamma _0}\left( {t_q^ +  - t_0^ +  - {T_0}} \right)}}\left\| {\Delta _q^\theta } \right\|h\left( {t - t_q^ + } \right)}  = {\beta _{{\rm{max}}}}.
\end{eqnarray}	
	
If the statement 5.2 of Assumption 1 holds and $k \ge {\gamma _0}$, then $\forall q \in \mathbb{N} {\rm{\;}}\left\| {\Delta _q^\theta } \right\| \le {c_q}{e^{ - k\left( {t_q^ +  - t_0^ + } \right)}} \le {c_q}{e^{ - {\gamma _0}\left( {t_q^ +  - t_0^ + } \right)}}{\rm{,\;}}{c_q} > {c_{q + 1}}$, and, as a result, the following estimate holds even in case of the infinite number of switches:
\begin{eqnarray}\label{eq_e30}
	\beta \left( t \right) \le \left\| {\tilde \theta \left( {t_0^ +  + {T_0}} \right)} \right\| + \sum\limits_{q = 1}^i {{c_q}{e^{ - {\gamma _0}{T_0}}}h\left( {t - t_q^ + } \right)}  \le \left\| {\tilde \theta \left( {t_0^ +  + {T_0}} \right)} \right\| + \sum\limits_{q = 1}^i {{c_q}h\left( {t - t_q^ + } \right)} {\rm{.}}	
\end{eqnarray}

The series in \eqref{eq_e30} is of positive terms, and all its partial sums are bounded by virtue of the monotonicity of $0 < {c_{q + 1}} < {c_q}$. Hence $\sum\limits_{q = 1}^\infty  {{c_q}h\left( {t - t_q^ + } \right)}  < \infty $, which results in $\beta \left( t \right) \le {\beta _{{\rm{max}}}}$.

Considering \eqref{eq_e30} or \eqref{eq_e29}, owing to $\Omega \left( t \right) \ge {\Omega _{LB}} > 0{\rm{\;}}\forall t \ge t_0^ + {\rm{ + }}{T_0}{\rm{\;}}$ and using the obtained fact that $\beta \left( t \right)$ is bounded, the following upper bound is obtained for the solution of \eqref{eq_e28}:
\begin{eqnarray}\label{eq_e31}
	\left\| {\tilde \theta \left( t \right)} \right\| \le {\beta _{\max }}{e^{ - {\gamma _0}\left( {t - t_0^ +  - {T_0}} \right)}} + {\gamma _0}\Omega _{LB}^{ - 1}\int\limits_{t_0^ +  + {T_0}}^t {{e^{ - {\gamma _0}\left( {t - \tau } \right)}}\left\| {d\left( \tau  \right)} \right\|d\tau } .
\end{eqnarray}

Then to prove exponential convergence of the parameter error $\tilde \theta \left( t \right)$ it is necessary to show exponential convergence to zero of the second summand in \eqref{eq_e31} as well. To this end, below we obtain the upper bound of the disturbance $d\left( t \right)$. By virtue of the expressions from \eqref{eq22a}-\eqref{eq22c}, $d\left( t \right)$ is differentiated to obtain:
\begin{eqnarray}\label{eq_e32}
\begin{array}{l}
\dot d\left( t \right) = \dot {\cal Y}\left( t \right) - \dot \Omega \left( t \right)\theta \left( t \right) - \Omega \left( t \right)\dot \theta \left( t \right) = \\
 =  - k\left( {{\cal Y}\left( t \right) - \Upsilon \left( t \right)} \right) + k\left( t \right)\left( {\Omega \left( t \right) - \Delta \left( t \right)} \right)\theta \left( t \right) - \Omega \left( t \right)\dot \theta \left( t \right) = \\
 =  - kd\left( t \right) + k\varepsilon \left( t \right) - \Omega \left( t \right)\dot \theta \left( t \right){\rm{,\;}}d\left( {t_0^ + } \right) = 0.
\end{array}
\end{eqnarray}

Taking the results of Proposition 3 into consideration, the disturbance $\varepsilon \left( t \right)$ is redefined on the basis of the Heaviside function:
\begin{eqnarray}\label{eq_e33}
	\varepsilon \left( t \right) =  - adj\left\{ {\omega \left( t \right)} \right\}\int\limits_{t_{i - 1}^ + }^{t_i^ + } {{e^{ - \int\limits_{t_{i - 1}^ + }^\tau  {\sigma ds} }}\varphi \left( \tau  \right){\varphi ^{\rm{T}}}\left( \tau  \right)} {\rm{\;}}d\tau \sum\limits_{q = 1}^i {\Delta _q^\theta h\left( {\hat t_i^ +  - t} \right)h\left( {t - t_i^ + } \right)} ,
\end{eqnarray}
from which we have:
\begin{eqnarray}\label{eq_e34}
	\mathop {\lim }\limits_{{\Delta _{pr}} \to 0} \varepsilon \left( t \right) = \mathop {\lim }\limits_{{\Delta _{pr}} \to 0} \sum\limits_{q = 1}^i {h\left( {\hat t_i^ +  - t} \right)h\left( {t - t_i^ + } \right)}  = 0.
\end{eqnarray}

Then, when ${\Delta _{pr}} \to 0$, the component of the zero-state solution of \eqref{eq_e32}, which is caused by the influence of $\varepsilon \left( t \right)$, globally exponentially decreases, affects only the quality of transients of the parameter error $\tilde \theta \left( t \right)$ and can be excluded from consideration. Taking into account such simplification, which is justified when ${\Delta _{pr}} \to 0$, the equation \eqref{eq_e32} is rewritten as:
\begin{eqnarray}\label{eq_e35}
	\dot d\left( t \right) =  - kd\left( t \right) - \Omega \left( t \right)\dot \theta \left( t \right){\rm{,\;}}d\left( {t_0^ + } \right) = d\left( {t_0^ +  + {T_0}} \right) = 0,
\end{eqnarray}
where $d\left( {t_0^ + } \right) = d\left( {t_0^ +  + {T_0}} \right) = 0$ as, considering the statement 1 of Assumption 1, the switches of the regression parameters values are impossible over the time range $\left[ {t_0^ + {\rm{;\;}}t_0^ +  + {T_0}} \right]$, and $\forall t \in \left[ {t_0^ + {\rm{;\;}}t_0^ +  + {T_0}} \right]$ it holds that $\dot \theta \left( t \right) = 0$.

Applying the sifting property of the Dirac function and the inequality $k \ge {\gamma _0}$, the upper bound of the solution of the equation \eqref{eq_e35} is written:
\begin{eqnarray}\label{eq_e36}
\begin{array}{c}
d\left( t \right) =  - \int\limits_{t_0^ +  + {T_0}}^t {{e^{ - k\left( {t - \tau } \right)}}\Omega \left( \tau  \right)\sum\limits_{q = 1}^i {\Delta _q^\theta \delta \left( {\tau  - t_q^ + } \right)} } d\tau  =  - \sum\limits_{q = 1}^i {{e^{ - k\left( {t - t_q^ + } \right)}}\Omega \left( {t_q^ + } \right)\Delta _q^\theta h\left( {t - t_q^ + } \right)}  = \\
 = \left( { - \sum\limits_{q = 1}^i {{e^{{\gamma _0}\left( {t_q^ +  - t_0^ +  - {T_0}} \right)}}\Omega \left( {t_q^ + } \right)\Delta _q^\theta h\left( {t - t_q^ + } \right)} } \right){e^{ - {\gamma _0}\left( {t - t_0^ +  - {T_0}} \right)}} \le \left( { - \sum\limits_{q = 1}^i {{e^{k\left( {t_q^ +  - t_0^ +  - {T_0}} \right)}}\Omega \left( {t_q^ + } \right)\Delta _q^\theta h\left( {t - t_q^ + } \right)} } \right){e^{ - {\gamma _0}\left( {t - t_0^ +  - {T_0}} \right)}}.
\end{array}
\end{eqnarray}

It is important to note here that, according to Assumption 1, the regression parameters switches are impossible over the time range $\left[ {t_0^ + {\rm{;\;}}t_0^ +  + {T_0}} \right)$, which results into summation from $q = 1$ to $i$ in \eqref{eq_e36}.

If the statement 5.1 from Assumption 1 holds and the number of the regression parameters switches is finite: $i \le {i_{{\rm{max}}}} < \infty $, then, considering ${\Omega _{UB}} \ge \Omega \left( t \right){\rm{\;}}\forall t \ge t_0^ + {\rm{ + }}{T_0}{\rm{\;}}$, the following upper bound holds:
\begin{eqnarray}\label{eq_e37}
	\left\| {d\left( t \right)} \right\| \le \left( {\sum\limits_{q = 1}^{{i_{\max }}} {{e^{k\left( {t_q^ +  - t_0^ +  - {T_0}} \right)}}{\Omega _{{\mathop{\rm UB}\nolimits} }}\left\| {\Delta _q^\theta } \right\|h\left( {t - t_q^ + } \right)} } \right){e^{ - {\gamma _0}\left( {t - t_0^ +  - {T_0}} \right)}} = {d_{\max }}{e^{ - {\gamma _0}\left( {t - t_0^ +  - {T_0}} \right)}}	
\end{eqnarray}

If the statement 5.2 from Assumption 1 holds, then $\forall q \in \mathbb{N} {\rm{\;}}\left\| {\Delta _q^\theta } \right\| \le {c_q}{e^{ - k\left( {t_q^ +  - t_0^ + } \right)}}{\rm{,\;}}{c_q} > {c_{q + 1}}$, and, considering \linebreak ${\Omega _{UB}} \ge \Omega \left( t \right){\rm{\;}}\forall t \ge t_0^ + {\rm{ + }}{T_0}{\rm{\;}}$, it is obtained from \eqref{eq_e36} that:
\begin{eqnarray}\label{eq_e38}
	\left\| {d\left( t \right)} \right\| \le \left( {{\Omega _{UB}}\sum\limits_{q = 1}^i {{c_q}h\left( {t - t_q^ + } \right)} } \right){e^{ - {\gamma _0}\left( {t - t_0^ +  - {T_0}} \right)}}. 	
\end{eqnarray}

All partial sums of the series of positive terms in \eqref{eq_e38} are bounded, hence $\sum\limits_{q = 1}^i {{c_q}h\left( {t - t_q^ + } \right)}  < \infty $ and the following estimate holds even for the case of infinite number of switches:
\begin{eqnarray}\label{eq_e39}
	\left\| {d\left( t \right)} \right\| \le {d_{\max }}{e^{ - {\gamma _0}\left( {t - t_0^ +  - {T_0}} \right)}}.
\end{eqnarray}

The upper bound \eqref{eq_e37} or \eqref{eq_e39} is substituted into \eqref{eq_e31} to obtain:
\begin{eqnarray}\label{eq_e40}
\begin{array}{c}
\left\| {\tilde \theta \left( t \right)} \right\| \le {\beta _{\max }}{e^{ - {\gamma _0}\left( {t - t_0^ +  - {T_0}} \right)}} + {\gamma _0}{d_{\max }}\Omega _{LB}^{ - 1}\int\limits_{t_0^ +  + {T_0}}^t {{e^{ - {\gamma _0}\left( {t - \tau } \right)}}{e^{ - {\gamma _0}\left( {\tau  - t_0^ +  - {T_0}} \right)}}d\tau }  = \\
 = {\beta _{\max }}{e^{ - {\gamma _0}\left( {t - t_0^ +  - {T_0}} \right)}} + {\gamma _0}{d_{\max }}\Omega _{LB}^{ - 1}{e^{ - {\gamma _0}\left( {t - t_0^ +  - {T_0}} \right)}}\int\limits_{t_0^ +  + {T_0}}^t {1d\tau }  = {\beta _{\max }}{e^{ - {\gamma _0}\left( {t - t_0^ +  - {T_0}} \right)}} + {\gamma _0}{d_{\max }}\Omega _{LB}^{ - 1}{e^{ - {\textstyle{{{\gamma _0}} \over 2}}\left( {t - t_0^ +  - {T_0}} \right)}}\chi \left( t \right),
\end{array}
\end{eqnarray}
where $\chi \left( t \right) = {e^{{\textstyle{{ - {\gamma _0}} \over 2}}\left( {t - t_0^ +  - {T_0}} \right)}}\left( {t - t_0^ +  - {T_0}} \right){\rm{,\;}}\chi \left( {t_0^ +  + {T_0}} \right) = 0.$

Let the boundedness of the function $\chi \left( t \right)$ be proved. To this end, $\chi \left( t \right)$ is differentiated with respect to time:
\begin{eqnarray}\label{eq_e41}
	\dot \chi \left( t \right) =  - {\textstyle{{{\gamma _0}} \over 2}}{e^{ - {\textstyle{{{\gamma _0}} \over 2}}\left( {t - t_0^ +  - {T_0}} \right)}}\left( {t - t_0^ +  - {T_0}} \right) + {e^{ - {\textstyle{{{\gamma _0}} \over 2}}\left( {t - t_0^ +  - {T_0}} \right)}} =  - {\textstyle{{{\gamma _0}} \over 2}}\chi \left( t \right) + {e^{ - {\textstyle{{{\gamma _0}} \over 2}}\left( {t - t_0^ +  - {T_0}} \right)}}{\rm{,\;}}\chi \left( {t_0^ +  + {T_0}} \right) = 0.
\end{eqnarray}

The upper bound of the solution of the equation \eqref{eq_e41} is written as:
\begin{eqnarray}\label{eq_e42}
	\left| {\chi \left( t \right)} \right| \le \left| {\;\;\int\limits_{t_0^ +  + {T_0}}^t {{e^{ - {\textstyle{{{\gamma _0}} \over 2}}\left( {t - \tau } \right)}}{e^{ - {\textstyle{{{\gamma _0}} \over 2}}\left( {\tau  - t_0^ +  - {T_0}} \right)}}d\tau } } \right| \le \left| {\;\;\int\limits_{t_0^ +  + {T_0}}^t {{e^{ - {\textstyle{{{\gamma _0}} \over 2}}\left( {\tau  - t_0^ +  - {T_0}} \right)}}d\tau } } \right|{\rm{ = }}{\textstyle{2 \over {{\gamma _0}}}}{\rm{,\;}}\chi \left( {t_0^ +  + {T_0}} \right) = 0.
\end{eqnarray}

Then, after substitution of obtained upper bound from \eqref{eq_e42} into \eqref{eq_e40} it is obtained:
\begin{eqnarray}\label{eq_e43}
	\left\| {\tilde \theta \left( t \right)} \right\| \le {\beta _{\max }}{e^{ - {\gamma _0}\left( {t - t_0^ +  - {T_0}} \right)}} + 2{d_{\max }}\Omega _{LB}^{ - 1}{e^{ - {\textstyle{{{\gamma _0}} \over 2}}\left( {t - t_0^ +  - {T_0}} \right)}} \le \left( {{\beta _{\max }} + 2{d_{\max }}\Omega _{LB}^{ - 1}} \right){e^{ - {\textstyle{{{\gamma _0}} \over 2}}\left( {t - t_0^ +  - {T_0}} \right)}}.
\end{eqnarray}

It follows from \eqref{eq_e43} that the parameter error $\tilde \theta \left( t \right)$ converges to zero exponentially with a rate being proportional to the parameter ${\gamma _0}$, which completes the proof of Theorem.

\nocite{*}
\bibliography{Glushchenko}%

\clearpage

\section*{Author Biography}

\begin{biography}
{\includegraphics[width=60pt,height=78pt]{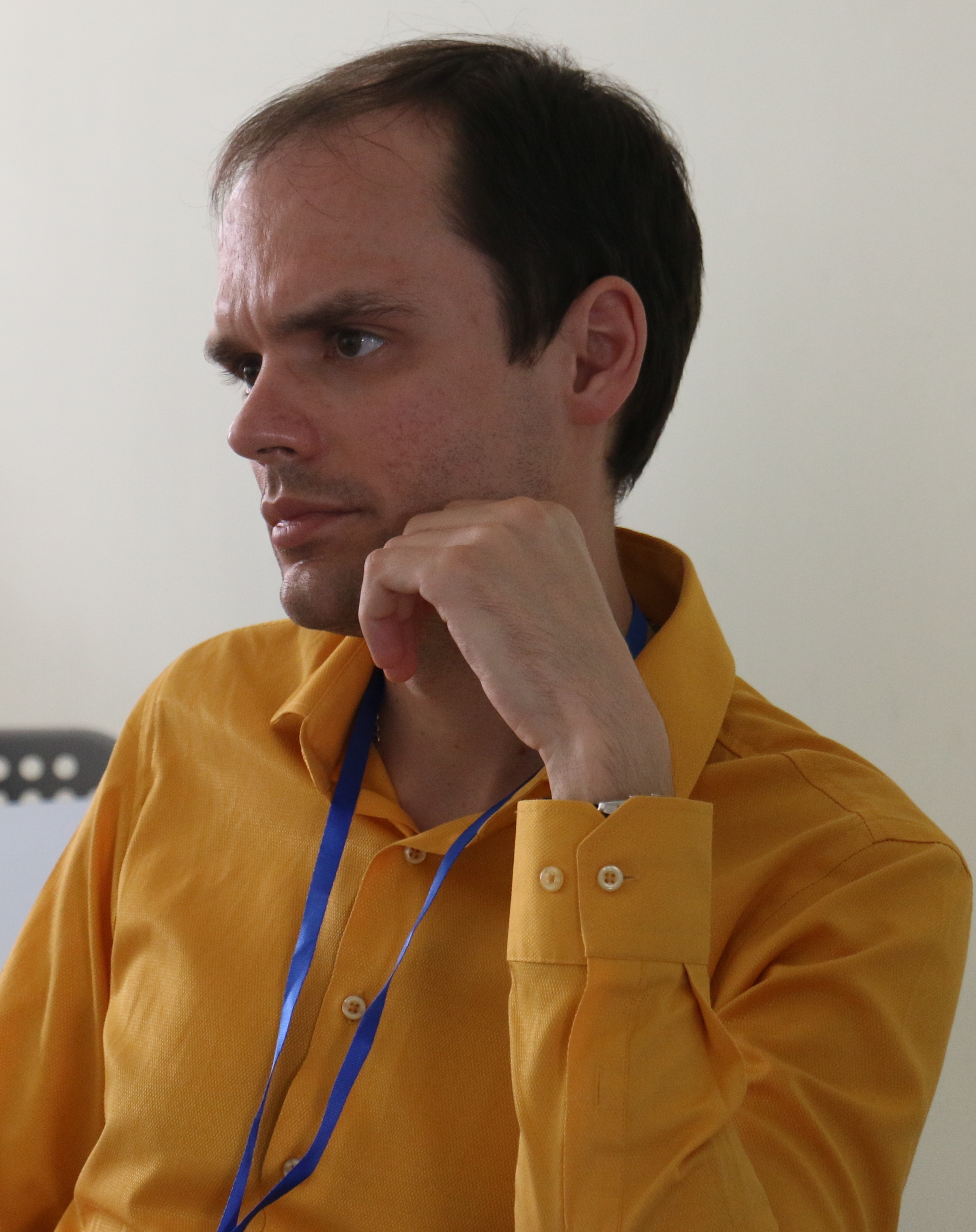}}{\textbf{Anton Glushchenko.} Anton Glushchenko received Software Engineering Degree from National University of Science and Technology "MISIS" (NUST "MISIS", Moscow, Russia) in 2008. In 2009 he gained Candidate of Sciences (Eng.) Degree from NUST "MISIS", in 2021 - Doctor of Sciences (Eng.) Degree from Voronezh State Technical University (Voronezh, Russia). Anton Glushchenko is currently the leading research scientist of laboratory 7 of V.A. Trapeznikov Institute of Control Sciences of Russian Academy of Sciences, Moscow, Russia. His research interests are mainly concentrated on exponentially stable adaptive control of linear time-invariant and time-varying plants under relaxed excitation conditions. The list of his works published includes more than 60 titles.}
\end{biography}

\begin{biography}
{\includegraphics[width=60pt,height=78pt]{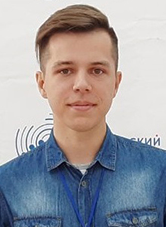}}{\textbf{Konstantin Lastochkin.} Konstantin Lastochkin received his bachelor's degree in Electrical Engineering from National University of Science and Technology "MISIS" (NUST "MISIS", Moscow, Russia) in 2020. The same year he enrolled Masters Degree course in automation of technological processes at NUST "MISIS". Konstantin Lastochkin is currently a research engineer of laboratory 7 of V.A. Trapeznikov Institute of Control Sciences of Russian Academy of Sciences, Moscow, Russia. His present interests include adaptive and robust control, identification theory, nonlinear systems. The list of his works published includes 20 titles.}
\end{biography}

\end{document}